\documentclass[manuscript=article,layout=twocolumn,10pt]{achemso}

\usepackage{graphicx}
\usepackage{amsmath}
\usepackage{xcolor}
\usepackage{dcolumn}
\usepackage{enumitem}
\usepackage{epstopdf}

\title{Helical spin order in EuFe$_2$As$_2$ and EuRbFe$_4$As$_4$ single crystals}

\author{I.A.~Golovchanskiy$^*$}
\affiliation{Moscow Institute of Physics and Technology, State University, 9 Institutskiy per., Dolgoprudny, Moscow Region, 141700, Russia}
\alsoaffiliation{National University of Science and Technology MISIS, 4 Leninsky prosp., Moscow, 119049, Russia}
\alsoaffiliation{Dukhov Research Institute of Automatics (VNIIA), 127055 Moscow, Russia}

\author{I.V.~Shchetinin}
\affiliation{National University of Science and Technology MISIS, 4 Leninsky prosp., Moscow, 119049, Russia}

\author{K.S.~Pervakov}
\affiliation{Ginzburg Center for High Temperature Superconductivity and Quantum Materials, P.N. Lebedev Physical Institute of the RAS, 119991, Moscow, Russia}

\author{V.A.~Vlasenko}
\affiliation{Ginzburg Center for High Temperature Superconductivity and Quantum Materials, P.N. Lebedev Physical Institute of the RAS, 119991, Moscow, Russia}

\author{P.S.~Dzhumaev}
\affiliation{National Research Nuclear University MEPhI (Moscow Engineering Physics Institute), 31 Kashirskoye Shosse, 115409 Moscow, Russia}

\author{O.V.~Emelianova}
\affiliation{National Research Nuclear University MEPhI (Moscow Engineering Physics Institute), 31 Kashirskoye Shosse, 115409 Moscow, Russia}

\author{D.S.~Baranov}
\affiliation{Moscow Institute of Physics and Technology, State University, 9 Institutskiy per., Dolgoprudny, Moscow Region, 141700, Russia}
\alsoaffiliation{Dukhov Research Institute of Automatics (VNIIA), 127055 Moscow, Russia}
\alsoaffiliation{National Research Nuclear University MEPhI (Moscow Engineering Physics Institute), 31 Kashirskoye Shosse, 115409 Moscow, Russia}
\alsoaffiliation{Skobeltsyn Institute of Nuclear Physics, MSU, Moscow, 119991, Russia}

\author{A.~S.~Astrakhantseva}
\affiliation{Moscow Institute of Physics and Technology, State University, 9 Institutskiy per., Dolgoprudny, Moscow Region, 141700, Russia}

\author{V.M.~Pudalov}
\affiliation{Ginzburg Center for High Temperature Superconductivity and Quantum Materials, P.N. Lebedev Physical Institute of the RAS, 119991, Moscow, Russia}

\author{V.S.~Stolyarov}
\affiliation{Moscow Institute of Physics and Technology, State University, 9 Institutskiy per., Dolgoprudny, Moscow Region, 141700, Russia}
\alsoaffiliation{Dukhov Research Institute of Automatics (VNIIA), 127055 Moscow, Russia}

\begin{document}

\begin{abstract}
In this work, fabrication and characterization of magnetic properties of EuFe$_2$As$_2$ and EuRbFe$_4$As$_4$ single crystals is reported.
Magnetization measurements of samples with well defined thin film geometry and crystal orientation demonstrate a striking similarity in ferromagnetic properties of Eu subsystems in these two compounds.
Measurements with magnetic field applied along ab crystal planes reveal meta-magnetic transition in both compounds.
Numerical studies employing the $J_{z1}$ and $J_{z2}$ Heisenberg model suggest that the ground state of the magnetic order in Eu subsystem for both compounds is the helical spin order with the helical angle about $2\pi/5$, while the meta-magnetic transition is the helix-to-fan first order phase transition.
\end{abstract}

\maketitle

\section{Introduction}

The new class of ferromagnetic superconductors based on EuFe$_2$As$_2$ parent compound is a unique playground in material science and solid state physics \cite{Ren_PRL_102_137002,Cao_JPCM_23_464204,Jeevan_PRB_83_054511,Nandi_PRB_89_014512,Vlasenko_SUST_33_084009,Liu_PRB_93_214503,Liu_PRB_96_224510,Smylie_PRB_98_104503,Iida_PRB_100_014506,Ghigo_PRR_1_033110,Kim_PRB, Stolyarov_JPCL_11_9393} due to accessible temperatures for the coexistence of superconducting and ferromagnetic orders on atomic scales of crystal lattice.
Superconductivity in EuFe$_2$As$_2$-based ferromagnetic superconductors emerges by doping with phosphorus \cite{Ren_PRL_102_137002,Cao_JPCM_23_464204,Jeevan_PRB_83_054511,Nandi_PRB_89_014512}, or rubidium \cite{Liu_PRB_93_214503,Liu_PRB_96_224510,Smylie_PRB_98_104503}.
When it comes to the interplay between the superconductivity and the ferromagnetism, the main focus in studies of such materials is mostly on its superconducting properties and on the physical origin behind the presence of the superconductivity.
In case of EuFeAs-based ferromagnetic superconductors the coexistence is considered between magnetic ordering of Eu$^{2+}$ ions with $S=7/2$ and superconducting ordering of Fe-3d
electrons.

On the other side, the ferromagnetic properties of EuFe$_2$As$_2$-based ferromagnetic superconductors are no less interesting.
As was reported, the EuFe$_2$As$_2$ parent compound itself demonstrate some anti-ferromagnetic properties with the easy-plane magnetic anisotropy along the crystallographic ab planes \cite{Jiang_NJP_11_025007}.
By doping of EuFe$_2$As$_2$ with phosphorus EuFe$_2$(As$_{1-x}$P$_x$)$_2$ ferromagnetic superconductor demonstrates ferromagnetic ordering with the easy-axis magnetic anisotropy along the c-axis (across crystallographic ab planes) \cite{Stolyarov_SciAdv_4_eaat1061,Grebenchuk_PRB_102_144501}.
Coexistence of superconductivity and magnetism in EuFe$_2$(As$_{1-x}$P$_x$)$_2$ results in contraction of the out-of-plane ferromagnetic domains and promotes domain branching \cite{Grebenchuk_PRB_102_144501}. 
By doping of EuFe$_2$As$_2$ with rubidium EuRbFe$_4$As$_4$ ferromagnetic superconductor demonstrates helical magnetic spin ordering \cite{Iida_PRB_100_014506}, which is argued\cite{Koshelev_PRB_100_224503} to be related to coexistence between the superconductivity and ferromagnetism.

Quite a lot of controvercies can be found in different studies of the magnetic order in EuFe$_2$As$_2$ and EuRbFe$_4$As$_4$, as well as in reports of the coexistence between supercodnuctivity and magnetism in EuRbFe$_4$As$_4$.
For instance, in Ref.~\cite{Jiang_NJP_11_025007} magnetization measurements of EuFe$_2$As$_2$ single crystal were reported, which reveal some metamagnetic transitions. 
These transitions will be confirmed in this work.
On the other side, in Ref.~\cite{Xiao_PRB_80_174424} neutron scattering measurements revealed A-type antiferromagnetic ordering in EuFe$_2$As$_2$, which is not subjected to metamagnetic transitions upon magnetization.
In Ref.~\cite{Collomb_PRL_126_157001} exchange interaction between the superconducting and magnetic subsystems was demonstrated with the scanning Hall microscopy.
In contrast, ARPES and EPR measurements in Refs.~\cite{Kim_PRB,Hemmida_PRB_103_195112} demonstrate the decoupling of the Eu$^{2+}$ magnetic moments from superconducting FeAs layers.
Moreover, the helical ordering reported in Ref.~\cite{Iida_PRB_100_014506} is in some contradiction with the basic $J_{z1}-J_{z2}$ Heisenberg model for helical spin systems \cite{Nagamiya_SSP_20_305,Robinson_PRB_2_2642,Johnston_PRL_109_077201,Johnston_PRB_91_064427}, since it makes undefined interactions between the nearest neighbor Eu-layers.

With this work we aim to toss some extra firewood into these debates.
We report a study of the spin order in EuFe$_2$As$_2$ and EuRbFe$_4$As$_4$ by measuring magnetization of single crystalline samples with well defined shape and crystallographic orientation.
We find that far below the Curie temperature ferromagnetic properties of Eu in both compounds EuFe$_2$As$_2$ and EuRbFe$_4$As$_4$ are practically identical despite the difference in composition and the presence of superconducting ordering in the later compound.
We observe the step-like magnetization of Eu subsystem in both EuFe$_2$As$_2$ and EuRbFe$_4$As$_4$, which we attribute to the helical spin ordering with the helical angle about $2\pi/5$ by numerical simulations.
We also determine that the c-crystal axis can not be associated with the hard magnetization axis.
We expect these results bring additional insight into the subject of the magnetic order in Eu subsystem in antiferromagneic EuFe$_2$As$_2$ and EuRbFe$_4$As$_4$ ferromagnetic superconductor.

\section{Fabrication and characterization of samples}

EuRbFe$_4$As$_4$ single crystal samples were grown using the self-flux method, by analogy with previous works \cite{Pervakov_SUST_26_015008,Vlasenko_SUST_33_084009}. 
The initial high purity components of Eu (99.95\%), Rb (99.99\%) and preliminary synthesized precursor FeAs (99.98\% Fe+99.9999\% As) were mixed with 1:1:12 molar ratio. 
The mixture was placed in an alumina crucible and sealed in a niobium tube under 0.2 atm of residual argon pressure. 
The sealed container was loaded into a tube furnace with an argon atmosphere to suppress the alkali metal evaporation. Next, the furnace was heated up to 1250$^\circ$C, held at this temperature for 24~h to homogenize melting, and then cooled down to 900$^\circ$C at a
rate of 2$^\circ$C/h. 
At this temperature, the ampoule with crystals was held for 24~h for growth defects elimination and then cooled down to room temperature inside the furnace. 
Finally, crystals were collected from the crucible in an argon glove box.
Optical microscopy reveals a well-defined layered structure of as-synthesized bulk crystals, and their pliability for cleavage and exfoliation along the layering direction only.

Composition and morphology analysis was carried-out with a number of flake-shaped samples cleaved from different as-synthesized bulk crystals 
by means of BSE SEM imaging, the energy-dispersive X-ray spectroscopy (EDX) and electron backscatter diffraction (EBSD) analysis.
Elemental analysis shows that synthesized crystals consist of EuFe$_2$As$_2$, EuRbFe$_4$As$_4$ phases and inclusions of Fe$_2$As.
Occasionally all three phases can be observed in the same flake sample.
EBSD data confirms crystal structure and orientation of observed phases.
Based on EBSD data the cleavage plane was always found to be \{001\} type.

%
\begin{figure}[!ht]
\begin{center}
\includegraphics[width=1\columnwidth]{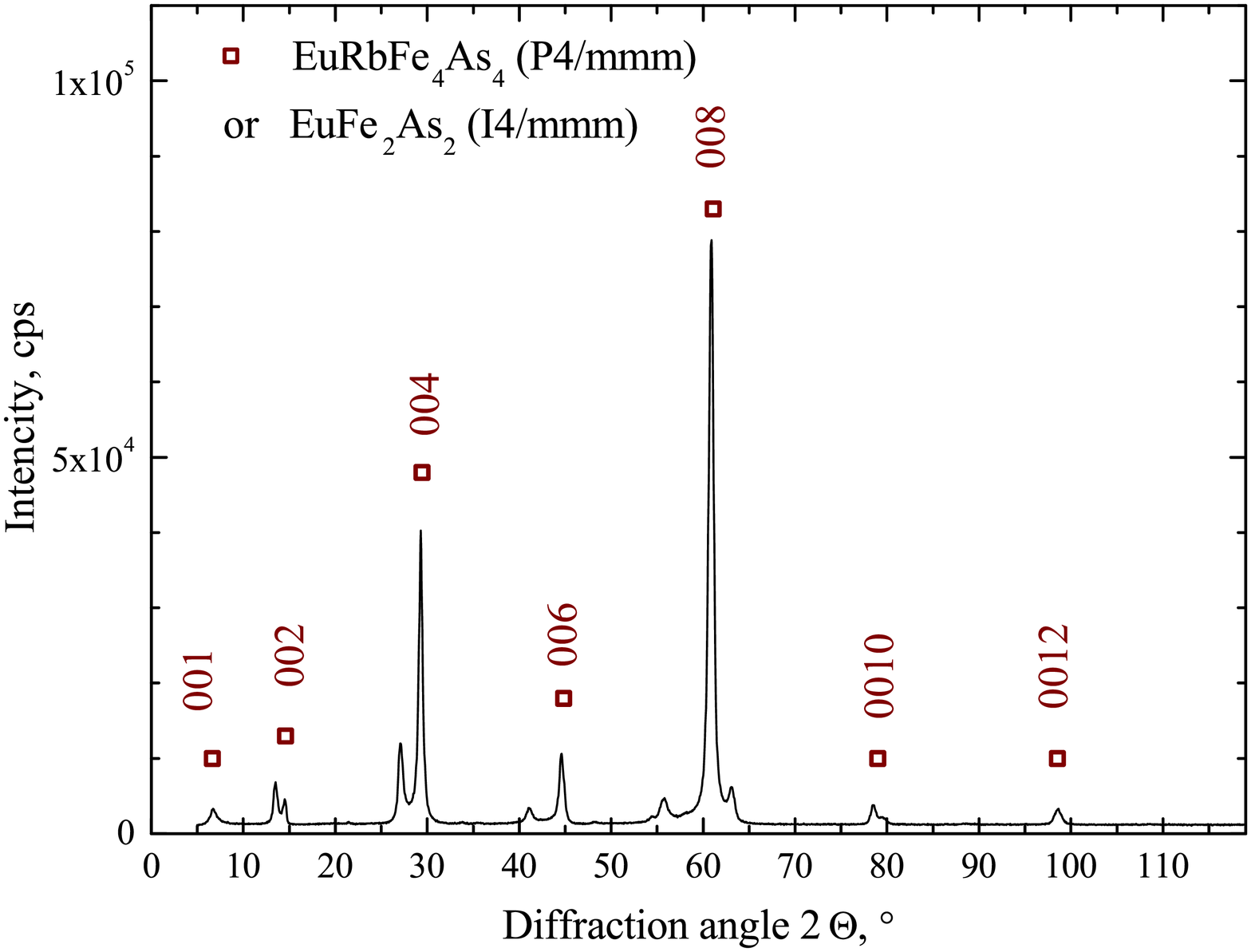}
\caption{Diffraction pattern from the SF4 flake sample (see details in supplementary S2 and S3).}
\label{XRD}
\end{center}
\end{figure}

Crystal structure and orientation of flake-shaped cleaved samples was studied with the X-ray diffraction analysis (XRD).
Figure~\ref{XRD} shows XRD spectrum of a sample of thickness about 30~$\mu$m.
The strongest diffraction lines are attributed to (001) type reflections of the EuFe$_2$As$_2$ or the EuRbFe$_4$As$_4$ phases, which indicates that the samples are single crystals oriented in the [001] direction along the normal to the plane of the flake. 
Importantly, the (001) reflection in Fig.~\ref{XRD} is attributed specifically to the EuRbFe$_4$As$_4$ phase \cite{Kawashima_JPSJ_85_064710}. 
The rest of peaks are attributed to the precursors used for the synthesis of these materials, as well as to oxidation products.
See details in supplementary S2.


%
\begin{figure}[!ht]
\begin{center}
\includegraphics[width=1\columnwidth]{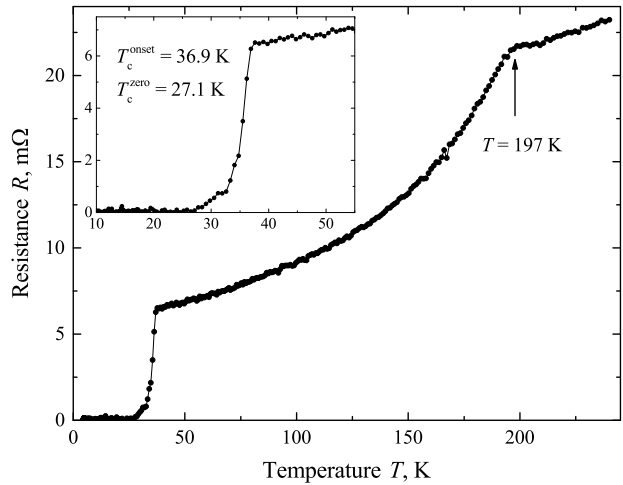}
\caption{The dependence of resistance of cleaved sample on temperature $R(T)$ at zero magnetic field.}
\label{RT}
\end{center}
\end{figure}

Electric transport properties were tested by studying resistance of cleaved samples at zero magnetic field using the pseudo-4-point measurement technique.
Figure~\ref{RT} shows the dependence of the electric ab-plane resistance $R$ on temperature.
The superconducting critical temperature $T^S_c =35.3$~K was determined by 50\% reducing of the resistance in the normal state before the drop (see the inset in Fig.~\ref{RT}).  
Upon cooling the onset of the superconducting transition occurs at the $T=36.9$~K, while the zero-resistance state is reached at $T=27.1$~K.

\begin{figure*}[!ht]
\begin{center}
\includegraphics[width=0.5\columnwidth]{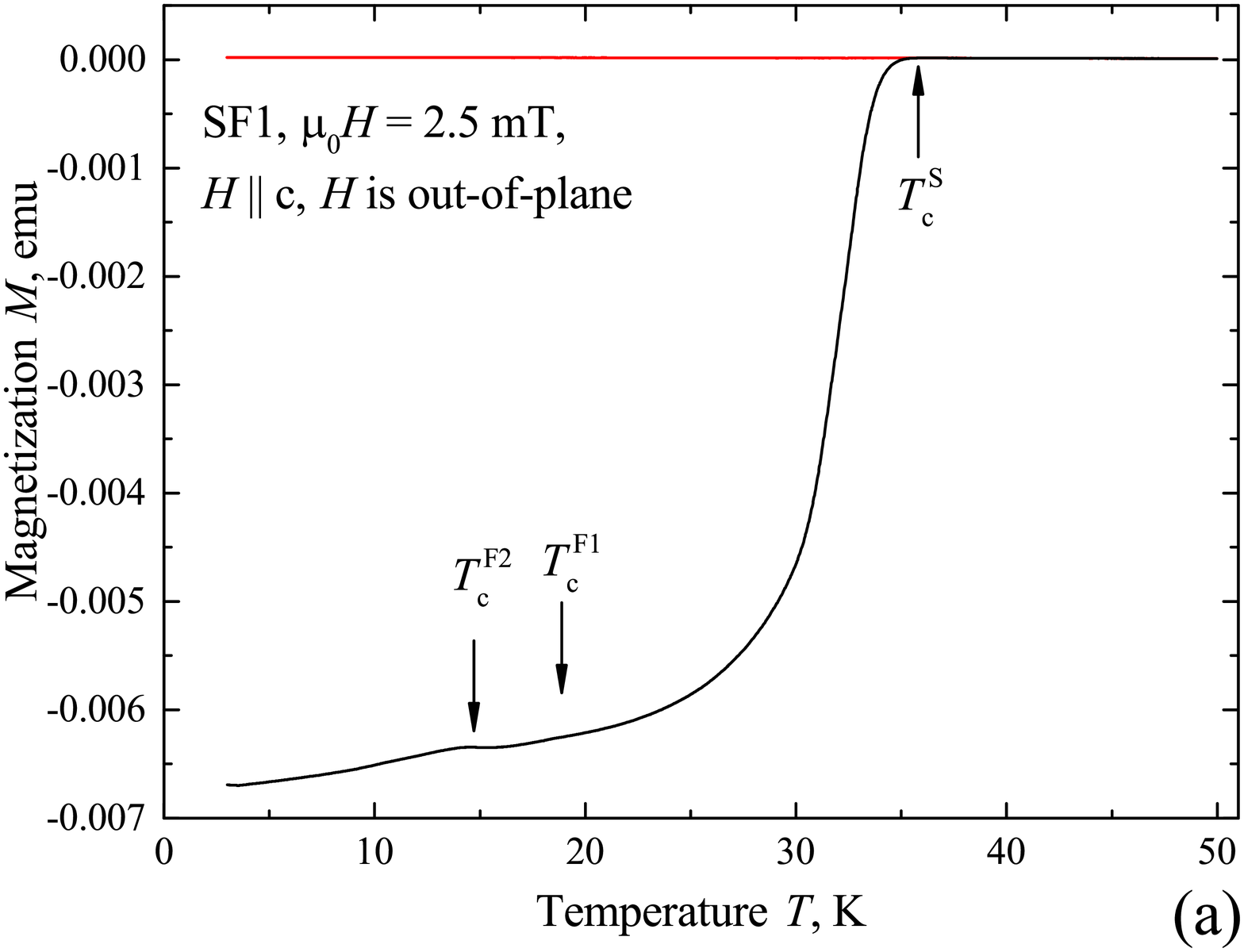}
\includegraphics[width=0.5\columnwidth]{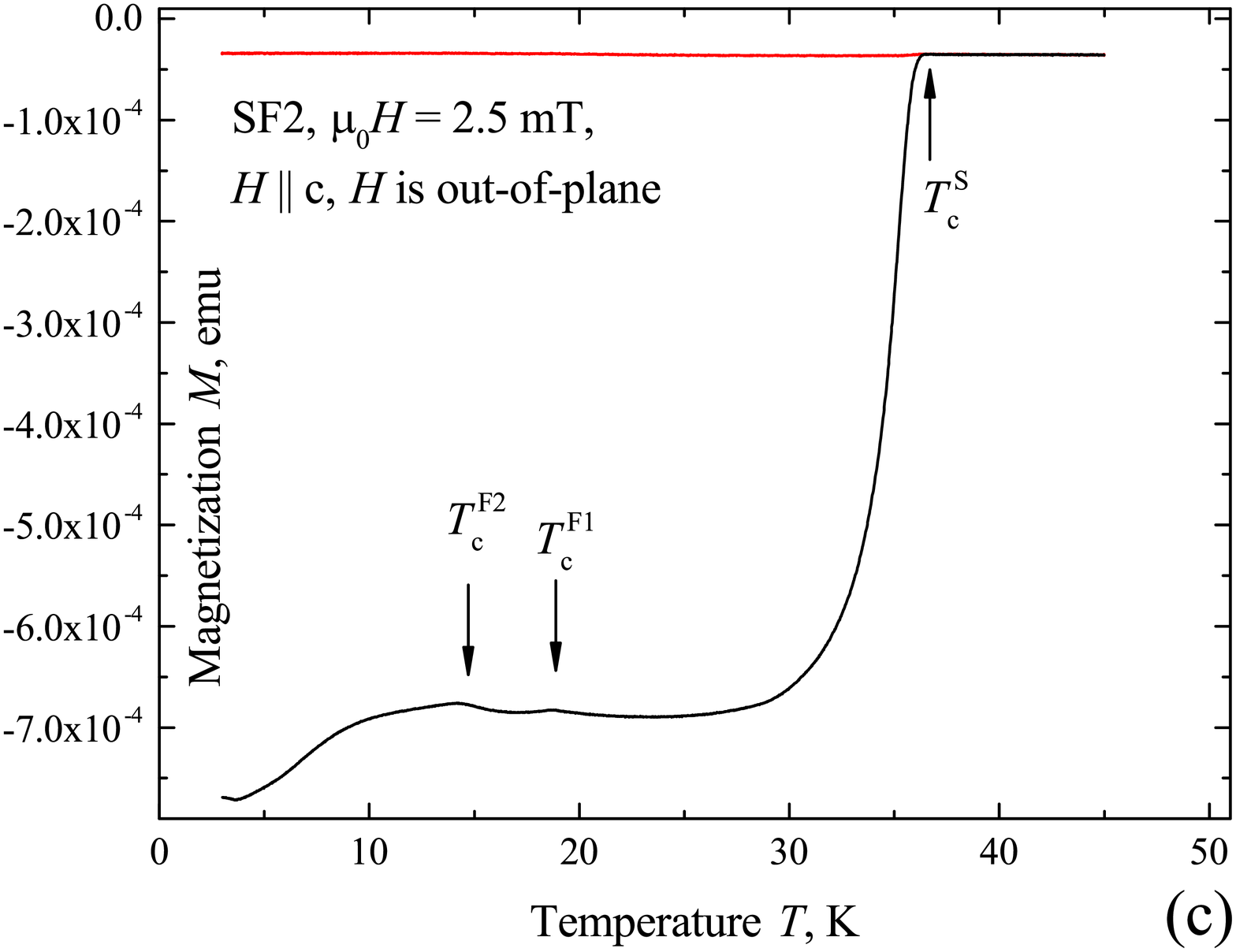}
\includegraphics[width=0.5\columnwidth]{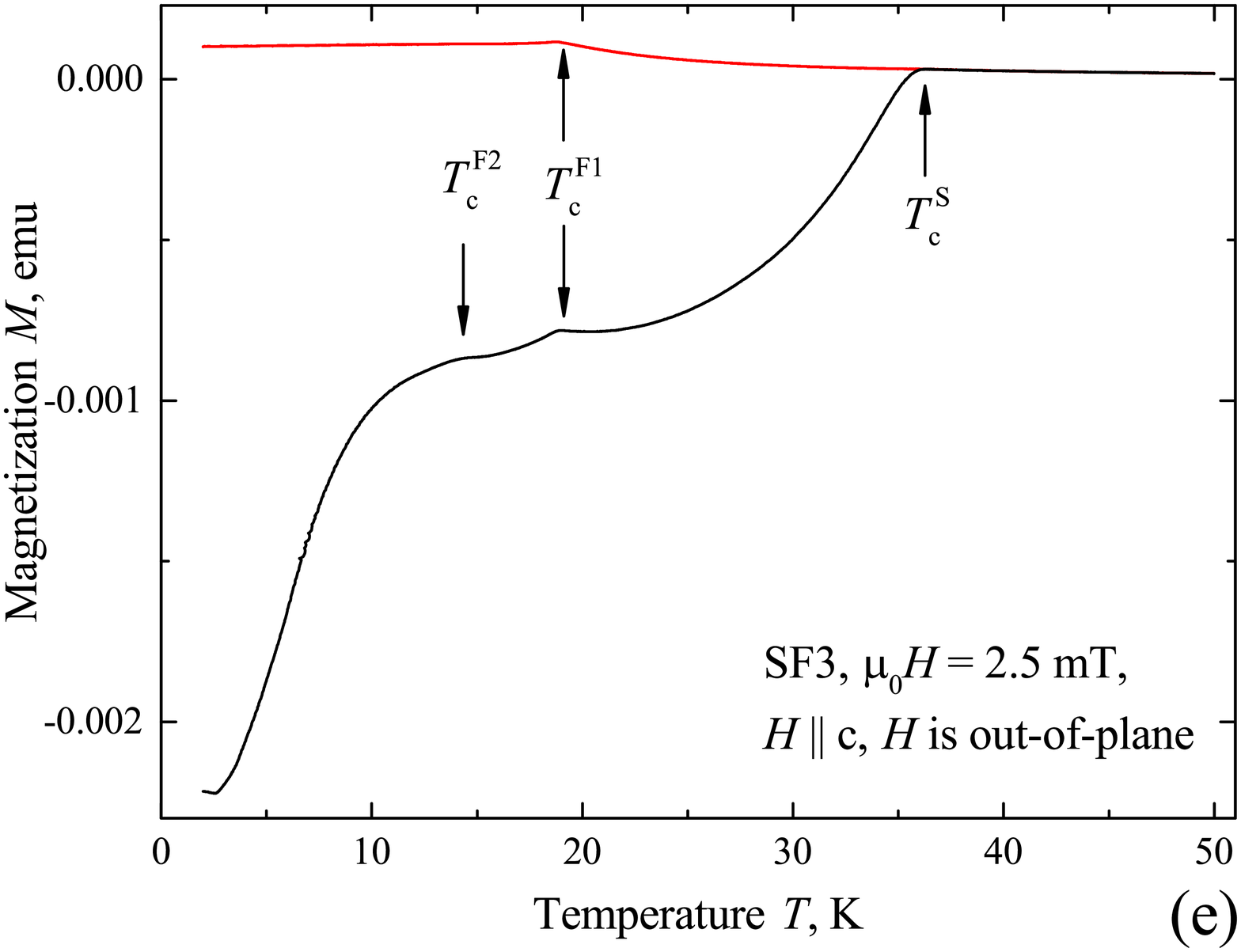} 
\\
\includegraphics[width=0.5\columnwidth]{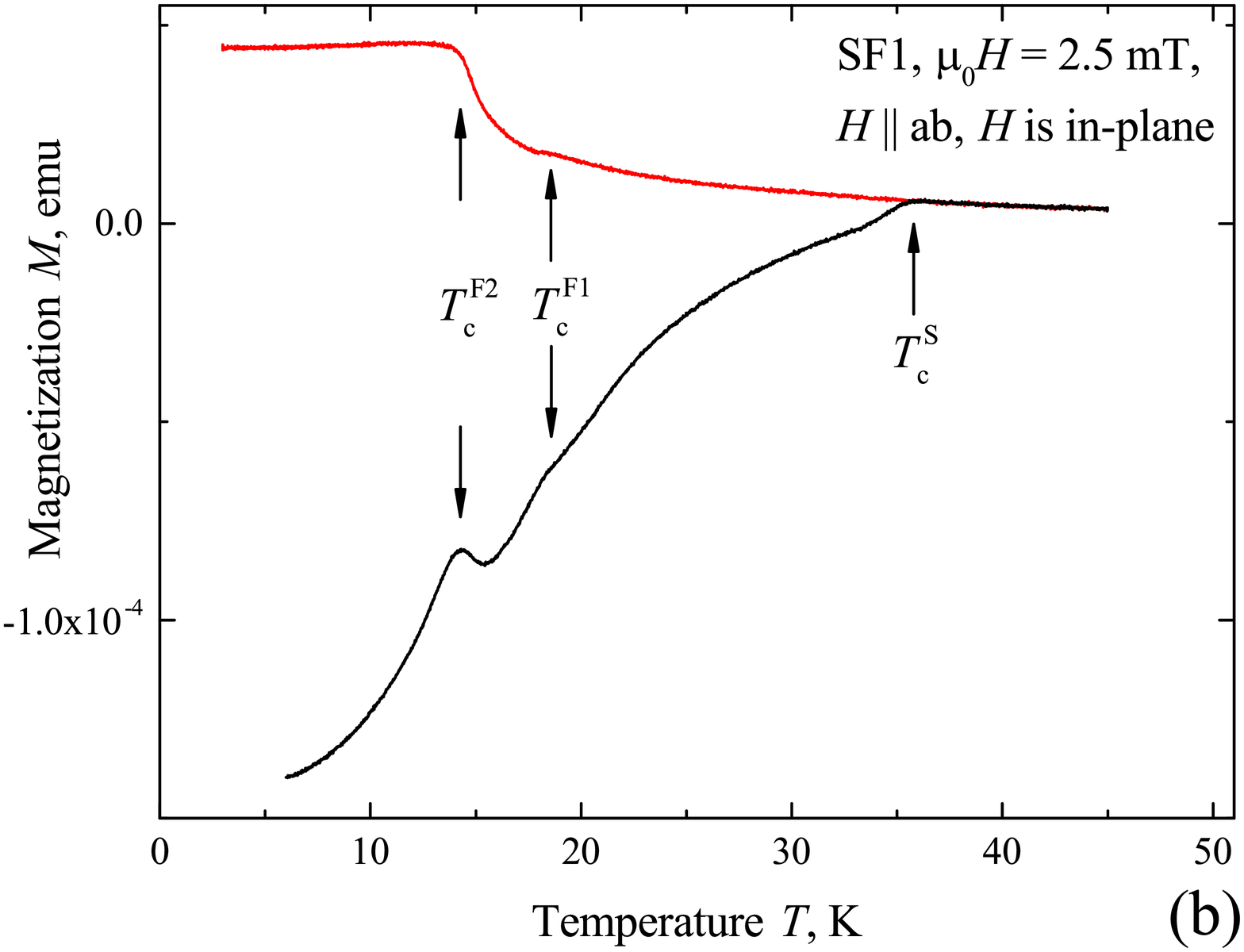}
\includegraphics[width=0.5\columnwidth]{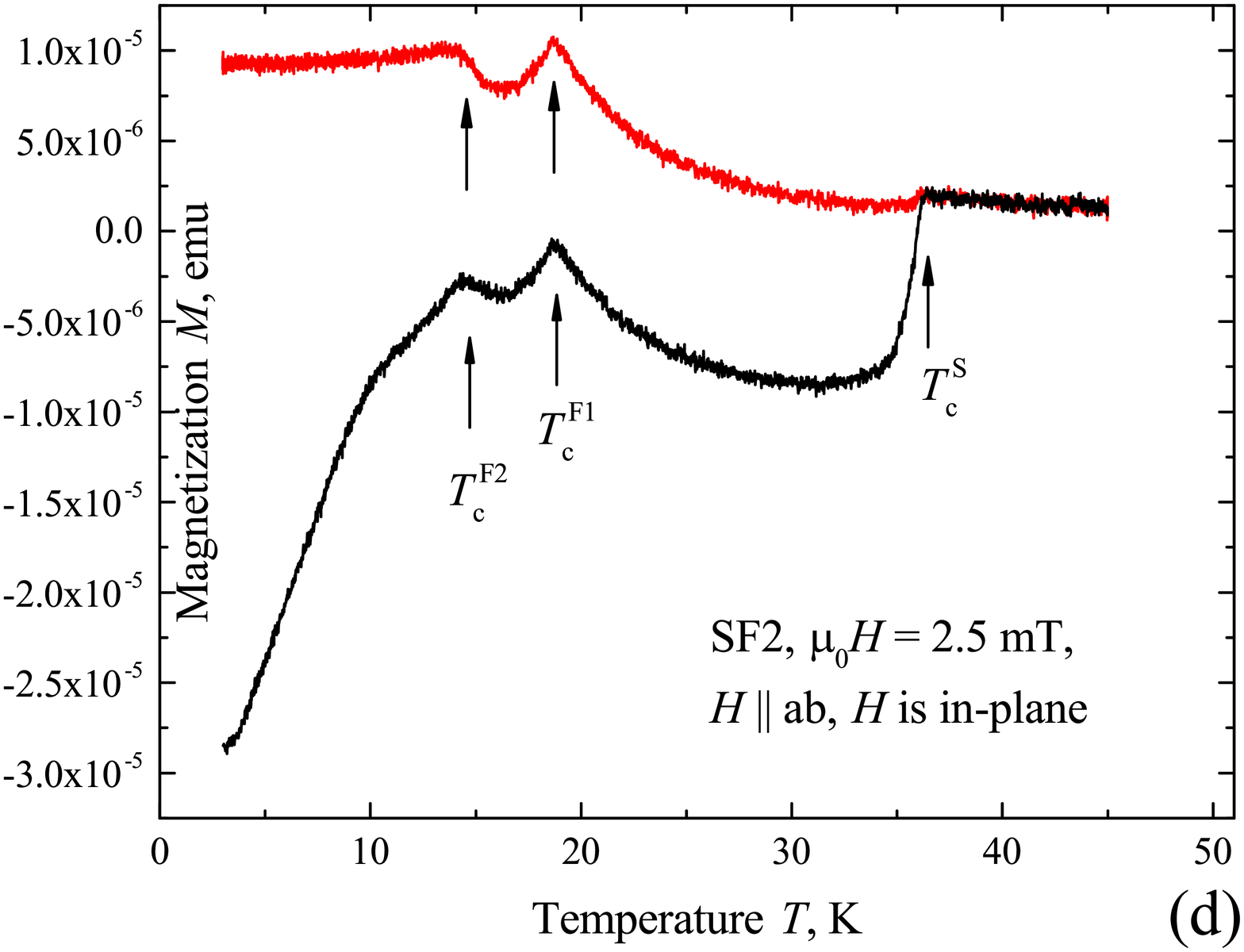}
\includegraphics[width=0.5\columnwidth]{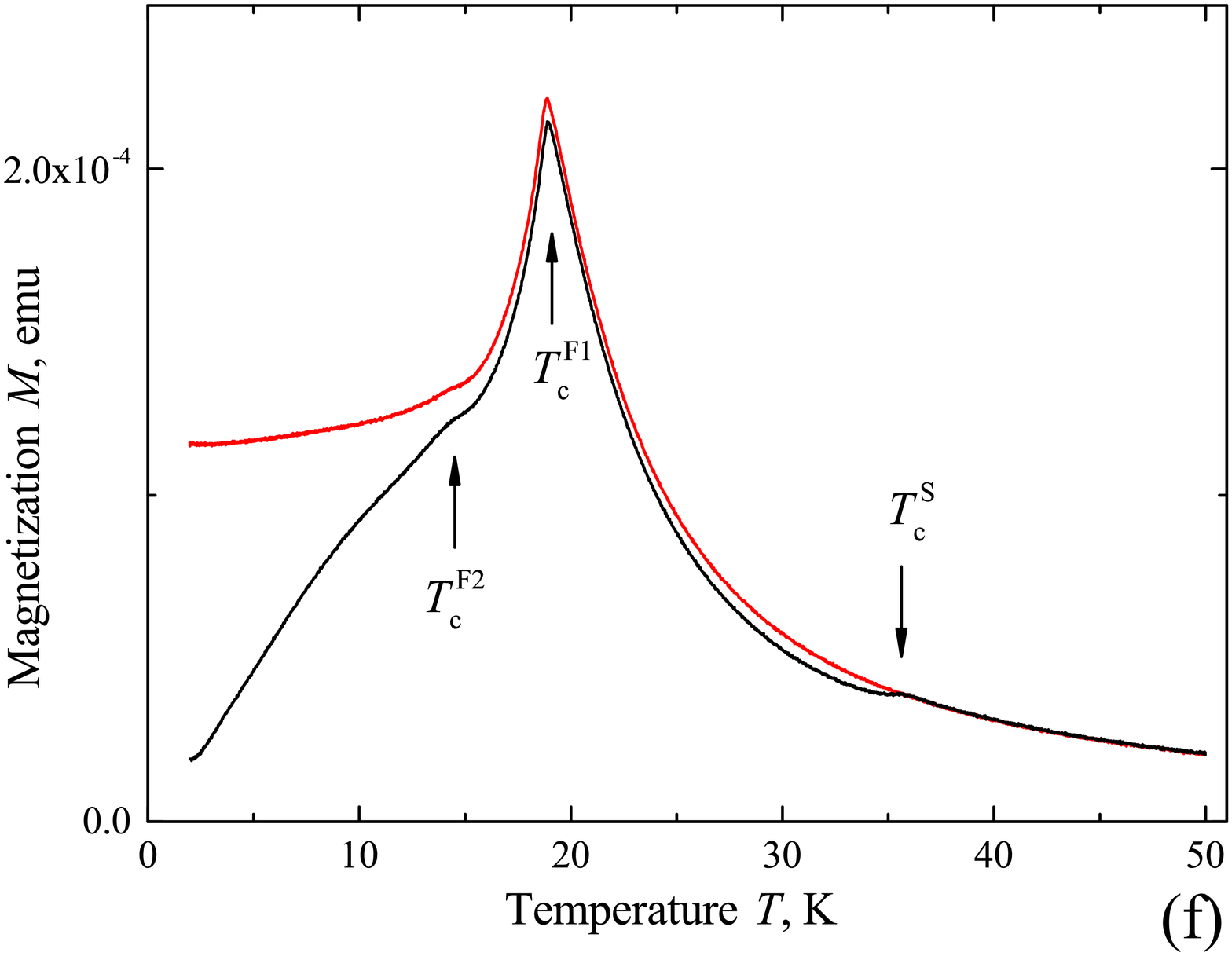}
\caption{The dependence of magnetization on temperature $M(T)$ for SF1 (a,b), SF2 (c,d), and SF3 (e,f) samples. 
Magnetization curves $M(T)$ are measured at $\mu_0 H=2.5$~mT in increasing temperature starting from the ZFC state (black curves) and in decreasing temperature starting the FC state (red curves).
a,c,e) $M(T)$ curves measured with the out-of-plane alignment of magnetic field along the $c$ crystal axis.
b,d,f) $M(T)$ curves measured with the in-plane alignment of magnetic field along the $ab$ crystal planes. 
Arrows indicate corresponding superconducting and ferromagnetic transition temperatures.  
}
\label{MT}
\end{center}
\end{figure*}

Overall, at $T>197$~K the sample demonstrates linear temperature dependence $R(T)$ attributed to metallic behavior. 
At $T<197$~K $R(T)$ curve changes its shape to the concave. 
Similar concave-shaped $R(T)$ behaviour was reported earlier for polycrystalline and single crystal EuFe$_2$As$_2$ \cite{Jeevan_PRB_78_052502,Ren_PRB_78_052501}.
For polycrystalline EuRbFe$_4$As$_4$ ferromagnetic superconductors an opposite convex-shaped temperature dependence was reported\cite{Liu_PRB_93_214503,Kawashima_JPSJ_85_064710}, which was associated with multiband effects with asymmetric scatterings in hole-doped iron-based superconductors\cite{Golubov_JETPLet_94_333}.
The transition at $T=197$~K was observed earlier in EuFe$_2$As$_2$, but not in EuRbFe$_4$As$_4$, and is attributed to the spin density wave (SDW)-type antiferromagnetic (AFM) transition due to itinerant Fe moments \cite{Jeevan_PRB_78_052502,Ren_PRB_78_052501}.
Therefore, diffused superconducting transition, concave-shaped $R(T)$, and the presence of anti-ferromagnetic transition in Fe indicate inhomogeneous phase composition of studied cleaved sample, which is consistent with composition analysis.

\section{Magnetization measurements}

Below we discuss three particular samples, indicated as SF1, SF2, and SF3, which in the most complete way reflect the range of magnetic properties observed for different samples  synthesized in this work. 
These samples were obtained by cleavage of as-synthesized crystals along the $c$-crystal axis.
Dimensions of SF1, SF2, and SF3 cleaved samples are up to a few mm in-plane along the $ab$ crystal planes, and about 20-50~$\mu$m in thickness along the $c$-axis.
These dimensions define the thin film measurement geometry for magnetization measurements.

\subsection{$M(T)$ curves}

Superconducting and ferromagnetic transition temperatures of synthesized samples were obtained by measuring dependencies of magnetization on temperature $M(T)$ starting from the zero-field-cooled (ZFC) and field-cooled (FC) states.
Figure~\ref{MT} shows $M(T)$ ZFC and FC curves for SF1, SF2, and SF3 samples measured at magnetic field 25~Oe applied out-of-plane along the $c$-axis (Fig.~\ref{MT}a,c,e) and in-plane along the $ab$-planes (Fig.~\ref{MT}b,d,f).
Superconducting transition temperature for SF1 sample $T^S_c=35.8$~K and of SF2 and SF3 samples $T^S_c=36.5$~K is defined at both orientation of the magnetic field as the irreversibility temperature between the ZFC and FC $M(T)$ curves.
This value is consistent with superconducting critical temperatures of EuRbFe$_4$As$_4$ reported previously \cite{Vlasenko_SUST_33_084009,Liu_PRB_93_214503,Liu_PRB_96_224510,Smylie_PRB_98_104503,Kawashima_JPSJ_85_064710}.

Ferromagnetic transition temperatures are determined as peaks or kinks on magnetization curves at $T<T^S_c$.
Two ferromagnetic transition temperatures $T^{F1}_c$ and $T^{F2}_c$ are observed for each sample.
For SF1 sample (Fig.~\ref{MT}a,b) the most pronounced is the ferromagnetic transition at $T^{F2}_c\approx14.5$~K, which corresponds to ferromagnetic ordering of Eu subsystem in EuRbFe$_4$As$_4$ ferromagnetic superconductor\cite{Vlasenko_SUST_33_084009,Liu_PRB_93_214503,Liu_PRB_96_224510,Kawashima_JPSJ_85_064710}.
In addition, a poorly-pronounced ferromagnetic transition is detected at $T^{F1}_c=18.6$~K, which corresponds to magnetic ordering temperature of Eu in EuFe$_2$As$_2$ parent compound\cite{Vlasenko_SUST_33_084009,Jiang_NJP_11_025007,Jeevan_PRB_78_052502,Ren_PRB_78_052501}.
Thus, based on $M(T)$ measurements we conclude that SF1 sample represents EuRbFe$_4$As$_4$ ferromagnetic superconductor with marginal content of EuFe$_2$As$_2$ phase.

For SF3 sample (Fig.~\ref{MT}e,f) intensity of magnetic transition peaks shows an opposite behaviour. 
The most pronounced is the ferromagnetic transition at $T^{F1}_c\approx18.9$~K, which corresponds to ferromagnetic ordering of Eu subsystem in EuFe$_2$As$_2$.
The transition of Eu subsystem in EuRbFe$_4$As$_4$ at $T^{F2}_c\approx14.3$~K is weakly pronounced.
Therefore, SF3 sample represents the EuFe$_2$As$_2$ parent compound with marginal content of EuRbFe$_4$As$_4$ ferromagnetic superconducting phase.
Moreover, $M(T)$ curves in in-plane geometry when magnetic field is applied along $ab$-planes (Fig.~\ref{MT}f) show a typical antiferromagnetic behaviour, which is consistent with previous reports\cite{Jiang_NJP_11_025007}.
SF2 sample (Fig.~\ref{MT}c,d) demonstrate an intermediate magnetic behaviour: magnetic transitions of Eu in both phases can be determined at $T^{F1}_c=18.6$~K and $T^{F2}_c=14.5$~K with comparable intensities at transition peaks.

Summing up, based on analysis of $M(T)$ magnetization curves, SF1 sample represents EuRbFe$_4$As$_4$ ferromagnetic superconductor with marginal content of EuFe$_2$As$_2$, SF3 sample represents the EuFe$_2$As$_2$ compound with small content of EuRbFe$_4$As$_4$ ferromagnetic superconducting phase, and SF2 sample represents some mixture of both EuFe$_2$As$_2$ and EuRbFe$_4$As$_4$ phases.
In other terms, all three samples do demonstrate superconductivity with the same $T^S_c=35.8\pm0.4$~K, show ZFC-FC hysteresis developed by the superconductivity, and reveal two ferromagnetic transitions $T^{F1}_c=18.7\pm0.2$~K $T^{F1}_c=14.4\pm0.1$~K attributed to Eu layers in EuFe$_2$As$_2$ and in EuRbFe$_4$As$_4$, respectively.
These conclusions are well supported by general behaviour of ZFC-FC magnetization curves (see supplementary S3 for an additional discussion).

\subsection{$M(H)$ curves}

\begin{figure*}[!ht]
\begin{center}
\includegraphics[width=0.5\columnwidth]{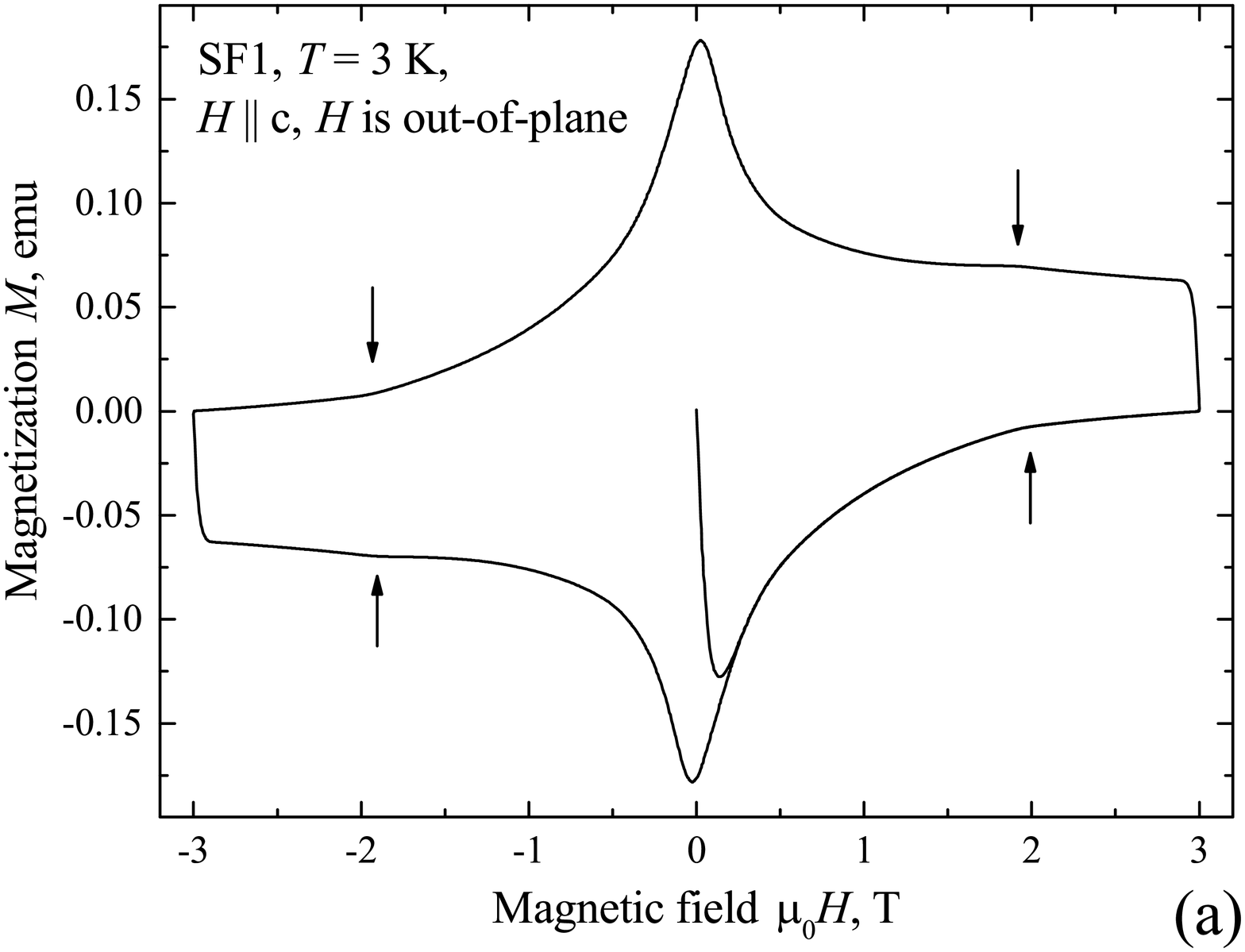}
\includegraphics[width=0.5\columnwidth]{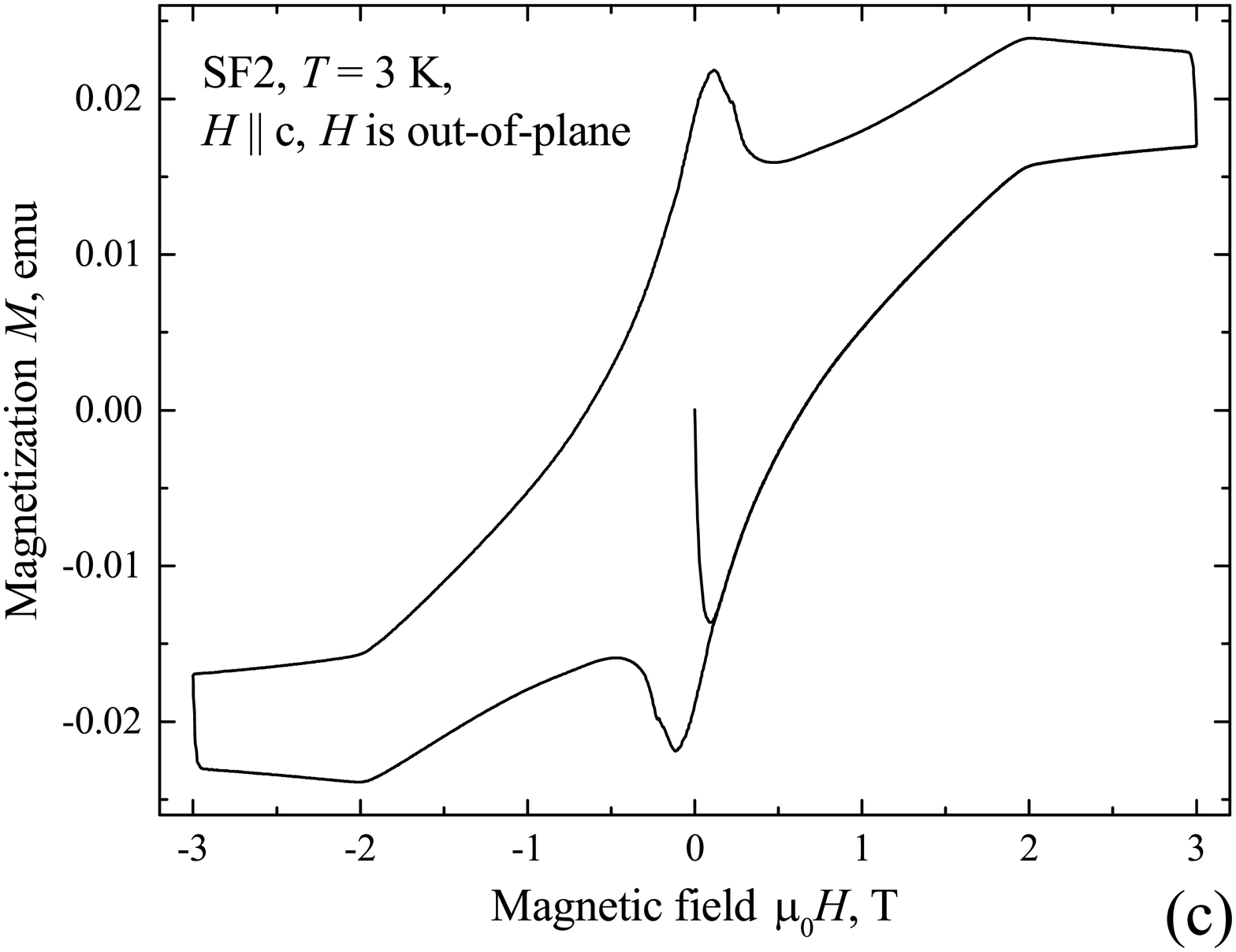}
\includegraphics[width=0.5\columnwidth]{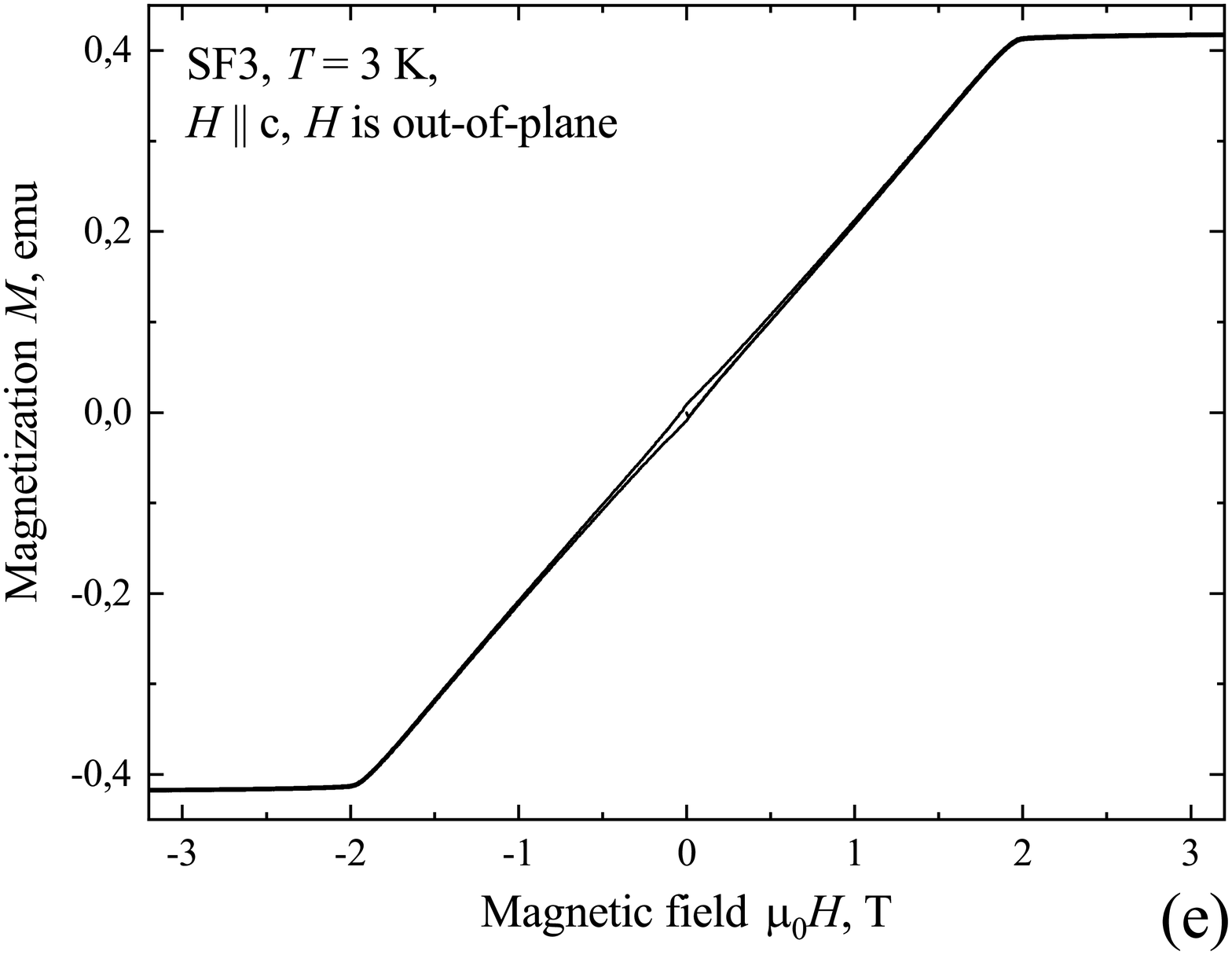}
\includegraphics[width=0.5\columnwidth]{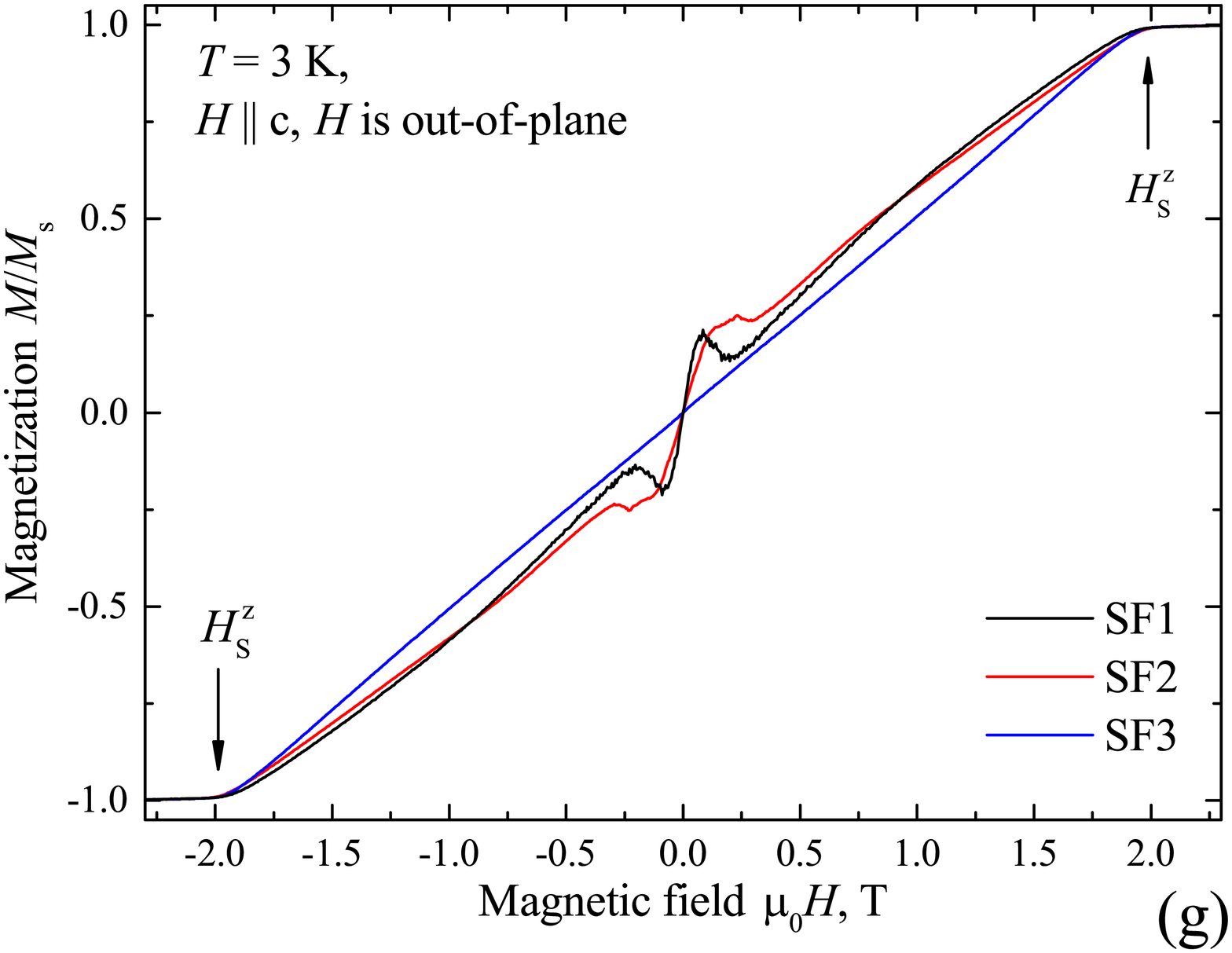}
\includegraphics[width=0.5\columnwidth]{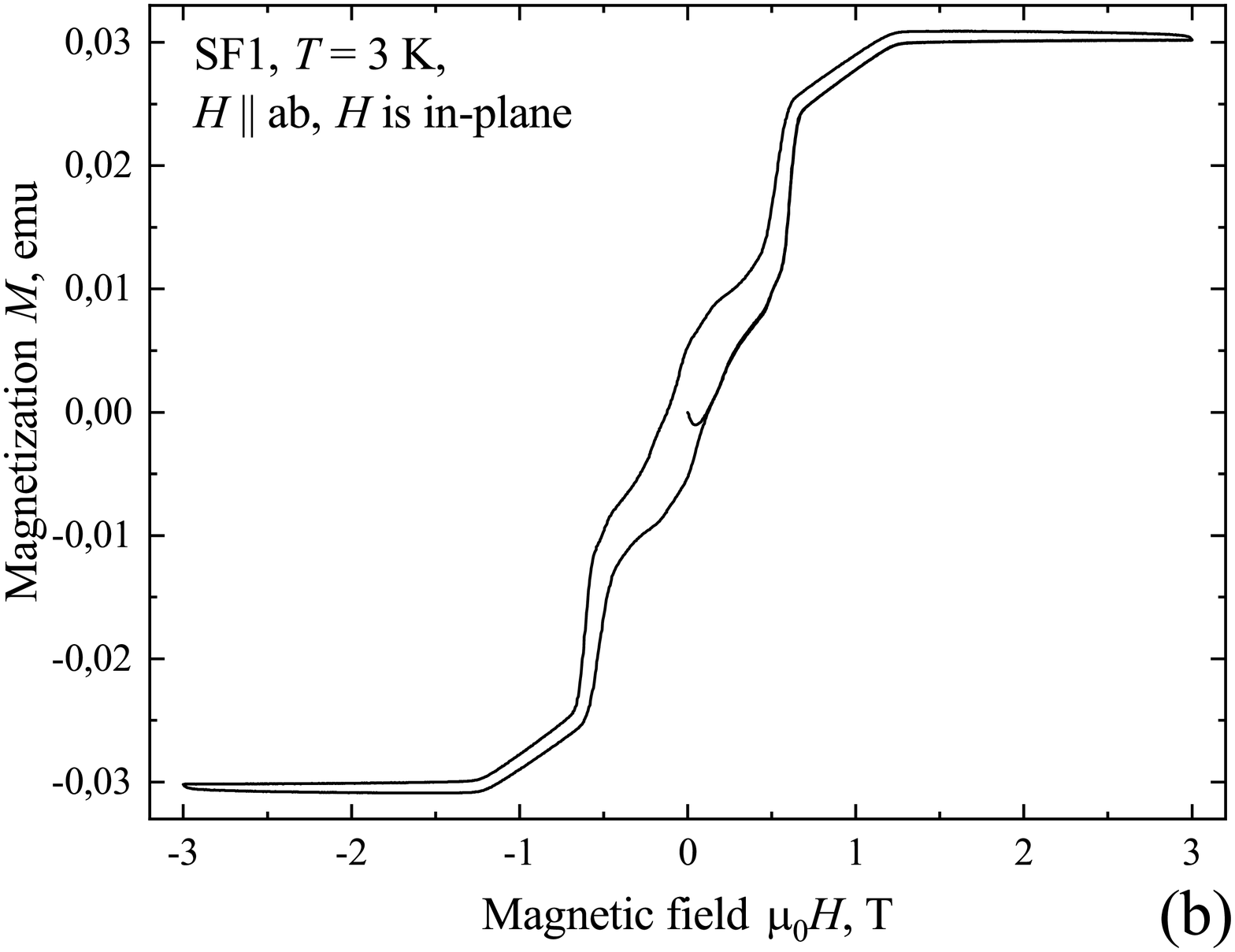}
\includegraphics[width=0.5\columnwidth]{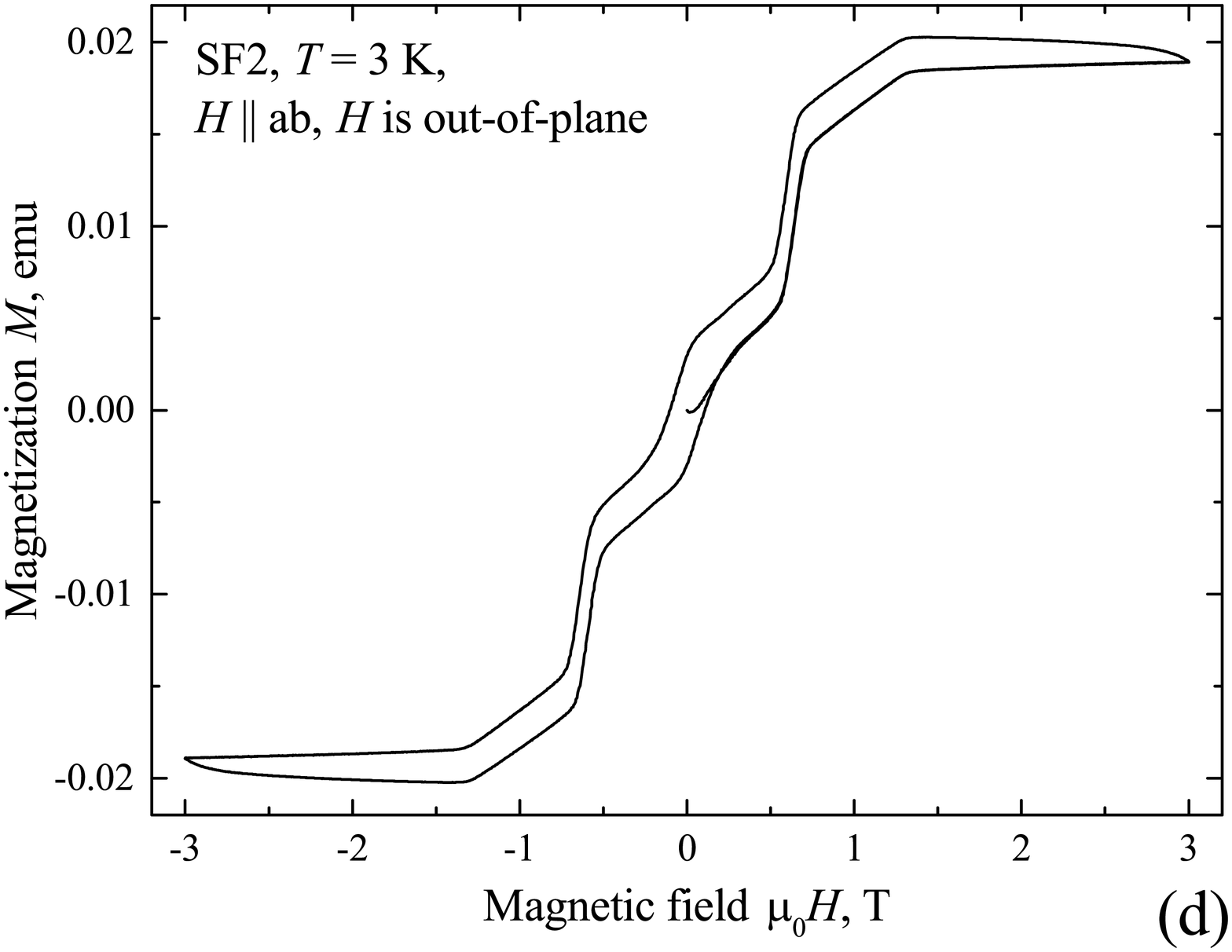}
\includegraphics[width=0.5\columnwidth]{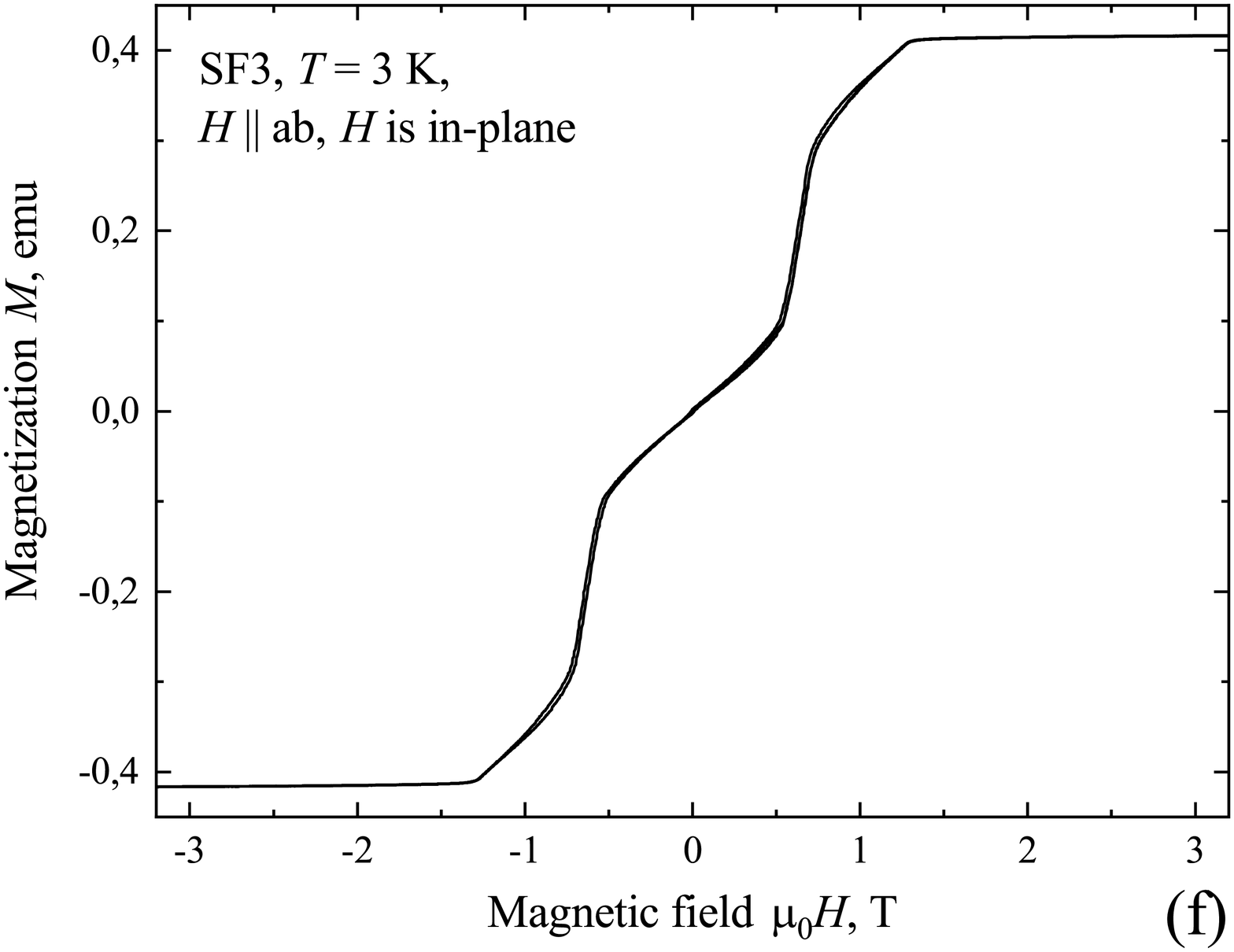}
\includegraphics[width=0.5\columnwidth]{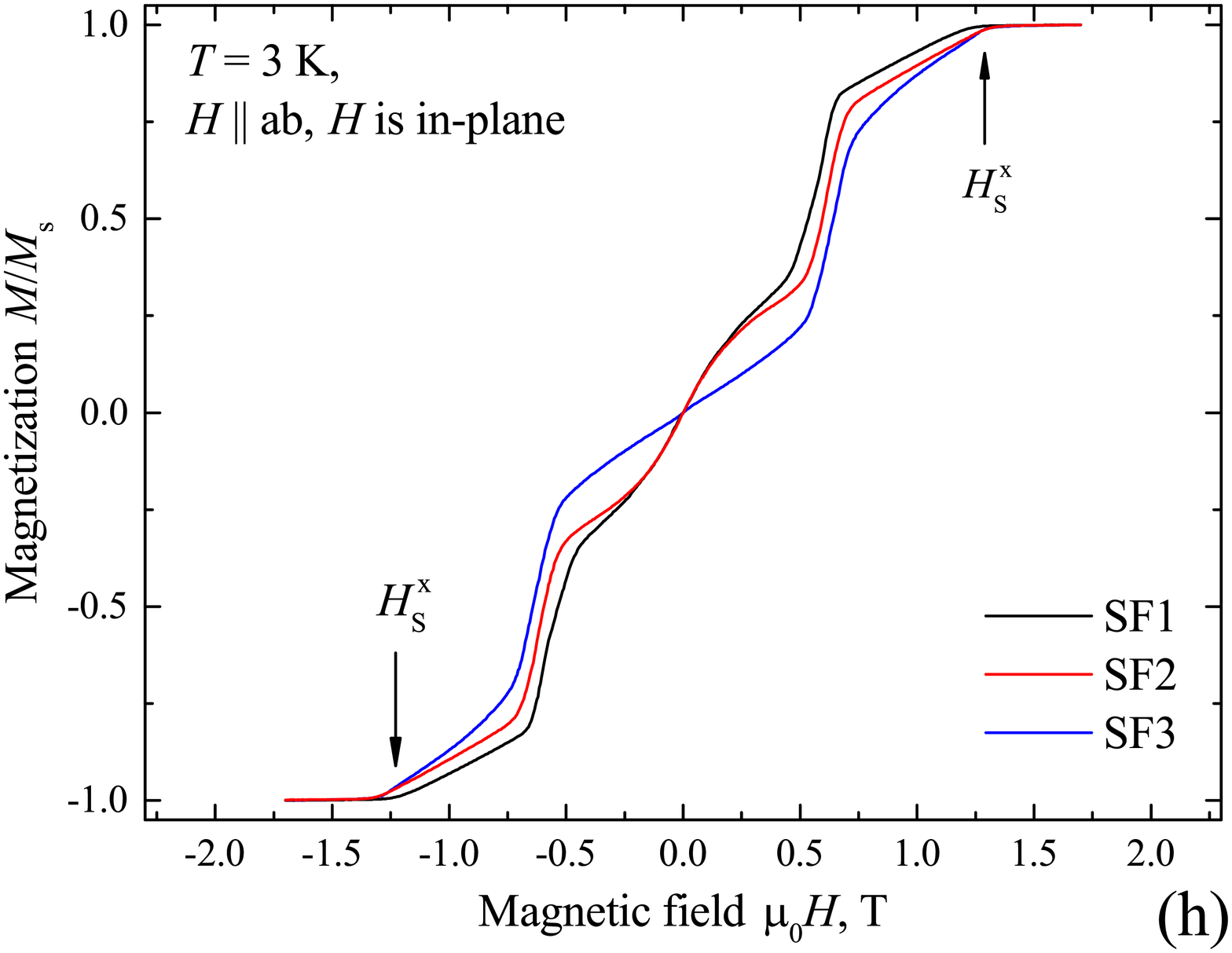}
\caption{
The dependence of magnetization on magnetic field $M(H)$ for SF1 (a,b), SF2 (c,d), and SF3 (e,f) samples.
Magnetization curves $M(H)$ are measured at $T=3$~K starting from the ZFC state.
a,c,e) $M(H)$ curves measured with the out-of-plane alignment of magnetic field along the $c$ crystal axis.
Arrows in (a) indicate kinks in $M(H)$ curves.
b,d,f) $M(H)$ curves measured with the in-plane alignment of magnetic field along the $ab$ crystal planes. 
g,h) Dependencies of magnetization of magnetic subsystem on magnetic field $M(H)$ obtained by averaging and normalization of raw $M(H)$ data in (a-f).
Arrows in (g,h) indicate ferromagnetic saturation fields $\mu_0 H^z_S \approx 1.97$~T and $\mu_0 H^x_S \approx 1.25$~T.
}
\label{MH}
\end{center}
\end{figure*}

Figure~\ref{MH} shows $M(H)$ hysteresis loops measured for SF1, SF2, and SF3 samples at both in-plane and out-of-plane orientations of applied magnetic field.
In general, $M(H)$ curves for $H||c$ (Fig.~\ref{MH}a,c,e) witness different contributions of superconducting phases for different samples, which is consistent with conclusions deduced from Fig.~\ref{MT}.
Magnetization curves $M(H)$ for SF1 sample (Fig.~\ref{MH}a) show a developed hysteresis loop typical for type-II superconducting films at out-of-plane magnetic fields.
The origin of hysteresis behavior is strong pinning of superconducting vortices in accordance with the Bean or Kim critical state models \cite{BEAN_RMP_36_31,Norris_JPDAP_3_489,Chen_JAP_66_2489}.
At the same time, ferromagnetic contribution of Eu subsystem manifests itself in anti-symmetry of the loop in respect to $H$-axis, as well as a minor kinks at $\mu_0 H\approx 2$~T.
Magnetization curves $M(H)$ for SF2 sample (Fig.~\ref{MH}b) show a modified hysteresis behaviour, which can be characterized as an over-layer of superconducting hysteresis loop and ferromagnetic magnetization curves of comparable magnetizations.
A more pronounced contribution of ferromagnetic subsystem can be explained by a presence of non-superconducting EuFe$_2$As$_2$ phase, in accordance with Fig.~\ref{MT}c,d.
The well-developed kinks at $\mu_0 H=\mu_0 H^z_S \approx 2$~T indicates the saturation field of ferromagnetic subsystem.
Magnetization curves $M(H)$ for SF3 sample (Fig.~\ref{MH}c) show a hysteresis-free loops typical for ferromagnetic thin films at out-of-plane magnetic field with the same saturation field about $\mu_0 H^z_S \approx 2$~T. 
Absence of the hysteresis indicates absence of contribution from superconducting EuRbFe$_4$As$_4$ phase, in accordance with Fig.~\ref{MT}e,f.

Measurements of $M(H)$ curves with $H||ab$ (Fig.~\ref{MH}b,d,f), in general, confirm contributions of superconducting phase into magnetization for different samples.
The strongest contribution of superconductivity (the strongest hysteresis) is observed for SF1 and SF2 samples, while SF3 sample shows hysteresis-free behaviour.  
In addition, measurements with $H||ab$ reveal step-like behaviour with three transition fields for every sample.
The highest-field transition at $\mu_0 H=\mu_0 H^x_S \approx 1.25$~T corresponds to the magnetic saturation of ferromagnetic Eu.
The step-like transitions at $H<H^x_S$ are attributed to metamagnetic phase transitions and are considered in details in the next section.

The main outcome of $M(H)$ magnetization measurements is summarized in Fig.~\ref{MH}g,h, where the contribution of the ferromagnetic subsystem into $M(H)$ magnetization is shown by subtracting the superconductivity-induced hysteresis.
Despite a demonstrated difference in phase compositions all three samples reveal identical properties of ferromagnetic subsystems, which are characterized by the same saturation fields at both out-of-plane and in-plane magnetic fields $\mu_0 H^z_S=1.97$ and $\mu_0 H^x_S=1.25$~T, as well as by rather close values of fields and magnetizations of metamagnetic transitions.
The first transition point is observed at the range $\mu_0 H\approx 0.45-0.55$~T and $M/M_s\approx 0.29-0.36$, the second transition point is observed at the range $\mu_0 H\approx 0.67-0.72$~T and $M/M_s\approx 0.78-0.85$.
Therefore, we are forced to conclude that \emph{ferromagnetic properties of Eu layers in EuRbFe$_4$As$_4$ and in EuFe$_2$As$_2$ are practically identical} at least at $T\ll T^{F1(F2)}_c$ and $T\ll T^S_c$.

It should be noted that in Fig.~\ref{MH}h a deviation of $M(H)$ from the linear dependence at $|\mu_0 H|<0.4$~T is observed for samples SF1 and SF2 .
Such deviation can be attributed to asymmetry in subtracted superconducting hysteresis loop, as well as to a response of ferromagnetic impurities, such as Fe$_2$As \cite{Hwang_JAP_105_07A946,Yang_PRB_102_064415} or FeAs (see supplementary S1 and S2).
For instance, the asymmetry of the subtracted superconducting hysteresis loop is manifested for SF1 and SF2 samples in Fig.~\ref{MH}g as curvatures at $|\mu_0 H|<0.3$~T.

\section{Helicity in E\lowercase{u}F\lowercase{e}$_2$A\lowercase{s}$_2$ and E\lowercase{u}R\lowercase{b}F\lowercase{e}$_4$A\lowercase{s}$_4$}

We state that the step-like magnetization in Fig.~\ref{MH}h is attributed to the non-collinear anti-ferromagnetic ordering of Eu-layers, known as the helical spin order.
Indeed, unconventional magnetic order of Eu was first noticed for EuFe$_2$As$_2$ parent compound \cite{Jiang_NJP_11_025007}.
In Ref. \cite{Jiang_NJP_11_025007} it was discussed that EuFe$_2$As$_2$ demonstrates easy-plane magnetization along ab planes, and at magnetic fields along the ab planes Eu layers undergo the spin-flop-like metamagnetic transition, which is different from the conventional anti-ferromagnetic spin-flop transition.  
Similar magnetic behaviour is shown in Fig.~\ref{MH}g,h.
Helical ordering of Eu in EuRbFe$_4$As$_4$ was observed recently with neutron scattering \cite{Iida_PRB_100_014506}.
Though, the helical angle $\phi_h=\pi/2$ between magnetic orientations in neighboring layers reported in Ref.~\cite{Iida_PRB_100_014506} is in some contradiction with the basic Heisenberg helical spin model due to the singularity at $\phi_h=\pi/2$ \cite{Nagamiya_SSP_20_305,Robinson_PRB_2_2642,Johnston_PRL_109_077201,Johnston_PRB_91_064427}.
Also, we note that a distinctive feature of the helical spin order with any acute helical angle  $\phi_h<\pi/2$ is the first-order helix-to-fan phase transition, which occurs at magnetic field of approximately the half of the saturation field\cite{Nagamiya_ProgTheorPhys_27_1253,Robinson_PRB_2_2642,Carazza_ZPhysB_84_301,Johnston_PRB_96_104405}. 
This transition is accompanied by the jump in magnetization.
In samples studied in this work (Fig.~\ref{MH}h) the metamagnetic transition occurs at $\mu_0 H\approx 0.5-0.7$~T, which is about a half of the saturation magnetization in ab-planes $\mu_0 H^x_S\approx1.25$~T.
At last, we notice that helicity is was confirmed for a variety of europium-based anti-ferromagnets \cite{Reehuis_JPhysChemSolids_53_687,Sangeetha_PRB_97_144403,Sangeetha_PRB_94_014422,Fabreges_PRB_93_214414}: EuCo$_2$P$_2$, EuCo$_2$As$_2$, and EuNiGe$_3$.
Thus, consideration of helical spin ordering in studied samples is fairly justified.

\subsection{Theoretical formalism and numerical details}

\begin{figure}[!ht]
\begin{center}
\includegraphics[width=0.7\columnwidth]{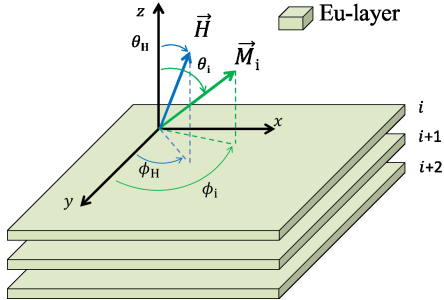}
\caption{Spherical coordinate system for orientation of magnetic moments in Eu layers $i$.
Coordinate z-axis is aligned with the c crystal axis.}
\label{coord}
\end{center}
\end{figure}

It appears that when the helical angle is not too small the configuration of the helical spin order can be deduced from the dependence of magnetization on magnetic field $M(H)$.
The dependence of the helical spin configuration and its magnetization on magnetic field has been considered extensively in a past by a number of groups analytically \cite{Enz_JAP_32_S22,Nagamiya_ProgTheorPhys_27_1253,Kitano_ProgTheorPhys_31_1,Nagamiya_SSP_20_305}, numerically \cite{Robinson_PRB_2_2642,Carazza_ZPhysB_84_301}, and recently within the molecular field theory \cite{Johnston_PRB_91_064427,Johnston_PRB_96_104405,Johnston_PRB_99_214438}.
In case of the planar layered anti-ferromagnetic arrangement the spin configuration can be obtained by employing the macrospin approximation for sub-lattice magnetizations and by minimizing the free energy of spin configurations.
Each macrospin characterizes coherent magnetization of an individual layer.
The total free energy of the macrospin configuration with $N$ macrospin Eu-layers is
\begin{equation}
\begin{aligned}
F={} & \sum_{i=1}^{N} [ \vec{M}_i\vec{H}+\frac{1}{2}\vec{M}_i\mathbf{\overline{N}}\vec{M}_i-K_u\cos^2(\theta_i)+\\
& +J_{z1}\vec{M}_i\vec{M}_{i+1}+J_{z2}\vec{M}_i\vec{M}_{i+2} ],
\end{aligned}
\label{F_en}
\end{equation}
where the first term is the Zeeman energy with the external field $\vec{H}$, the second term is the demagnetizing energy of individualr Eu-layers with $\mathbf{\overline{N}}$ being the demagnetizing factor, $K_u$ is the out-of-plane uniaxial magnetic anizotropy of the macrospin layer, $J_{z1}$ and $J_{z2}$ are the exchange interaction coefficients between the nearest neighbor and the next-nearest neighbor layers, respectively.
In spherical coordinates (the film sample is in x-y plane, $\theta$ and $\phi$ are the polar and the azimuthal angles, respectively, see Fig.~\ref{coord}) the general expression for energy of the spin configuration is reduced to
\begin{equation}
\begin{aligned}
f={} & F/M_s= \sum_{i=1}^{N} [ -H\cos(\alpha_{i,H})-\frac{1}{2}M_{eff}\cos^2(\theta_i)+\\
& +H_{z1}\cos(\alpha_{i,i+1})+H_{z2}\cos(\alpha_{i,i+2})],
\end{aligned}
\label{E_gen}
\end{equation}
where $M_s$ is the saturation magnetization of the Eu layer, $M_{eff}=M_s-2K_u/M_s$ is the effective saturation magnetization, which incorporates the out-of-plane anisotropy field, $H_{z1(z2)}=J_{z1(z2)}M_s$ are the corresponding exchange fields, $\cos{\alpha_{i,H}}$ and $\cos{\alpha_{i,j}}$ are cosines between the macrospin vector $i$ and external field, and between macrospin vectors $i$ and $j$, respectively. 
In spherical coordinates the cosine between two vectors is
\begin{equation}
\begin{aligned}
\cos(\alpha_{i,j})={} & \sin(\theta_i)\sin(\theta_j)\cos(\phi_i-\phi_j)+\\
& +\cos(\theta_i)\cos(\theta_j).
\end{aligned}
\label{alpha}
\end{equation}

Characteristic parameters of the spin system can be obtained by considering energy minima $\partial f/\partial \theta_i=\partial f/\partial \phi_i=0$ of particular spin configurations.
In case when the out-of-plane z-direction represents the hard axis, which is the case of this work, at zero field macrospin magnetizations are locked to x-y plane (a.k.a. 1D spin chains) and the ground state configuration for the spin-system depends on exchange field parameters $H_{z1}$ and $H_{z2}$.
In case when both $H_{z1}<0$ and $H_{z2}<0$ the ground state is ferromagnetic.
In case when $H_{z1}>0$ and $H_{z2}<0$ the ground state is anti-ferromagnetic.
Helical configurations correspond to $H_{z1}<0$ \& $H_{z2}>0$, or $H_{z1}>0$ \& $H_{z2}>0$.
From Eq.~\ref{E_gen} it follows that the energy of the helical spin configuration at zero field with helicity $N_h$ and the helical angle $\phi_h$ defined as $\phi_h=\phi_i-\phi_{i+1}=2\pi/N_h$ is 
\begin{equation}
f/N=H_{z1}\cos(\phi_h)+H_{z2}\cos(2\phi_h).
\label{E_hel}
\end{equation}
Minimization of Eq.~\ref{E_hel} in respect to the helical angle $\phi_h$ provides the equilibrium condition for exchange field parameters
\begin{equation}
H_{z1}+4H_{z2}\cos(\phi_h)=0.
\label{H_e}
\end{equation}
This relation implies that the stable helical spin configuration can be obtained when $|H_{z1}/4H_{z2}|<1$.

In case of the saturation of the spin system along the x-direction (in Fig.~\ref{coord} $\phi_H=0$ and $\theta_H=\pi/2$) minimization of the free energy of spin chain provides the saturation field \cite{Nagamiya_ProgTheorPhys_27_1253} 
\begin{equation}
H^x_S=2H_{z1}(1-\cos(\phi_h))+2H_{z2}(1-\cos(2\phi_h)).
\label{H_x}
\end{equation}
Notice that according to Eq.~\ref{H_e}~and~\ref{H_x} the saturation field $H^x_S$ defines both parameters of the exchange interaction $H_{z1}$ and $H_{z2}$ in their relation with the helicity $N_h$.

In case of the saturation of the spin system along the z-direction (in Fig.~\ref{coord} $\theta_H=0$) minimization of the free energy of spin chain provides the saturation field \cite{Kitano_ProgTheorPhys_31_1}
\begin{equation}
\begin{aligned}
H^z_S{} & =2H_{z1}(1-\cos(\phi_h))+ \\
& +2H_{z2}(1-\cos(2\phi_h))+M_{eff}.
\end{aligned}
\label{H_z}
\end{equation}

According to Eq.~\ref{H_x}~and~\ref{H_z} the difference in saturation fields provides the effective saturation magnetization $H^z_S-H^x_S=M_{eff}$, which recovers the result of the Stoner-Wohlfarth model.
According to Fig.~\ref{MH}g,h, at 3~K $\mu_0 M_{eff}=0.72$~T for all three samples.
Large effective magnetization secures in-plane magnetization in Eu layers when magnetic field is applied along $ab$-planes.
This alignment is critical for consideration of spin ordering as a one-dimensional spin chain, as done in Refs.~\cite{Nagamiya_ProgTheorPhys_27_1253,Robinson_PRB_2_2642,Carazza_ZPhysB_84_301,Johnston_PRB_96_104405}.

\begin{figure*}[!ht]
\begin{center}
\includegraphics[width=0.66\columnwidth]{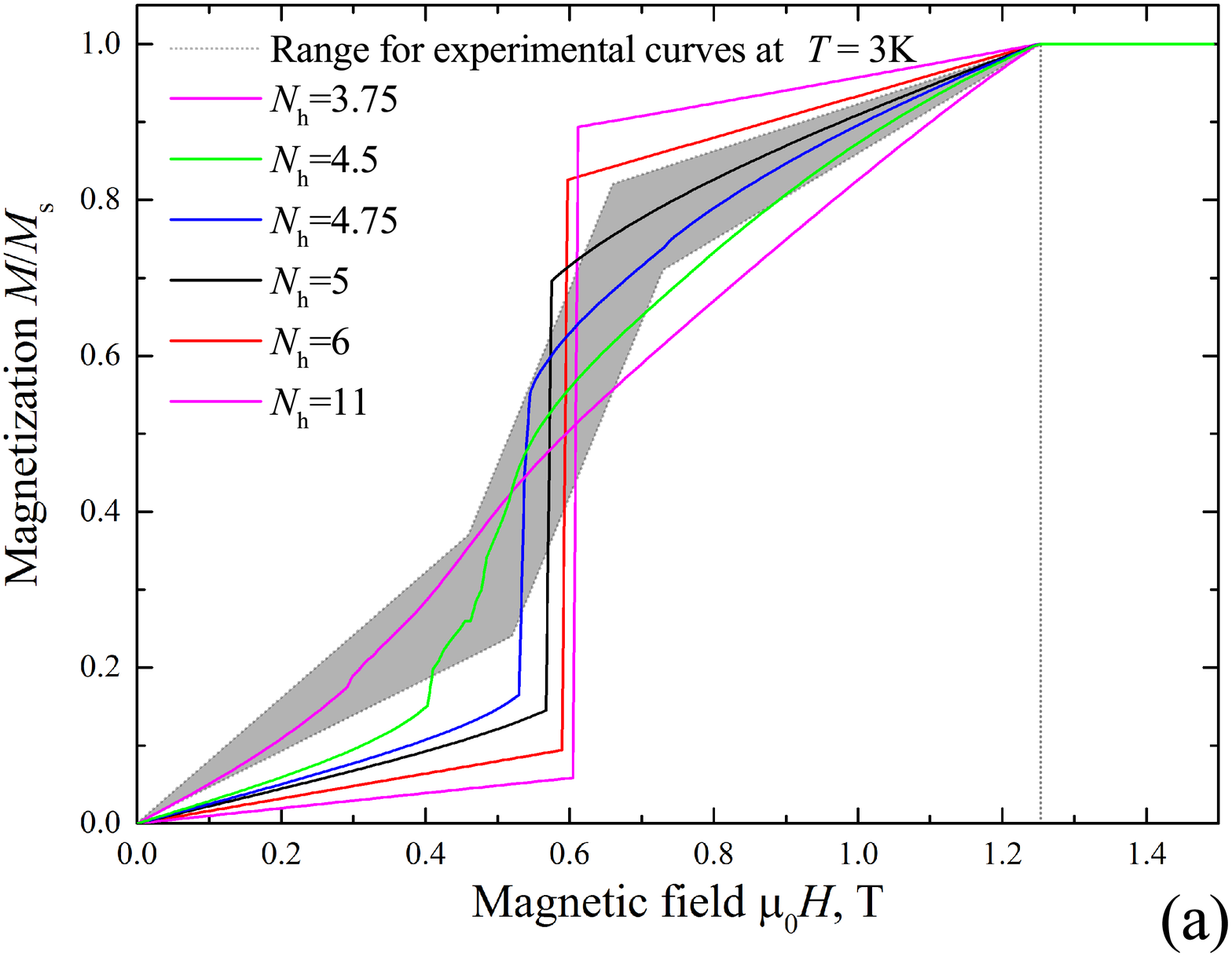}
\includegraphics[width=0.746\columnwidth]{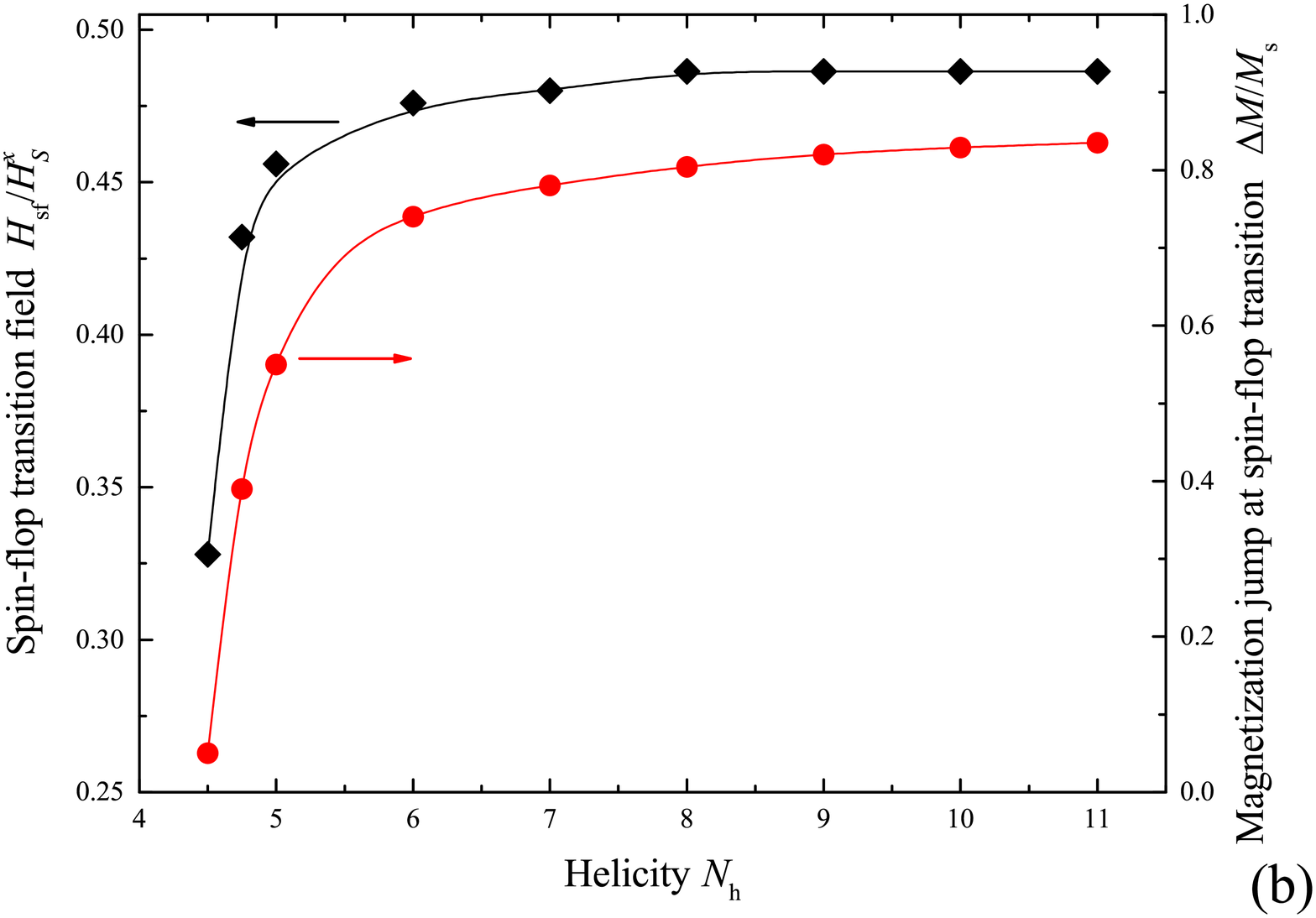}
\caption{
a) Theoretical dependencies $M(H)$ for helical magnets with different helicities $N_h$.
b) The dependence of the spin-flop field $H_{sf}/H^x_S$ (shown in black) and of the magnetization jump at spin-flop transition $\Delta M_{sf}/M_s$ (shown in red) on helicity $N_h$.
Lines in (b) are given as an eye-guide.
}
\label{Th}
\end{center}
\end{figure*}

The effective magnetization can be used for estimation of the magnetocrystalline anisotropy field $2K_u/M_s$.
Earlier it was shown that for both compounds EuRbFe$_4$As$_4$ and EuFe$_2$As$_2$  \cite{Ishida_npjQM_4_27,Liu_PRB_93_214503,Liu_PRB_96_224510,Smylie_PRB_98_104503} magnetization measurement suggests that Eu$^{2+}$ ions has local magnetic moment with $S = 7/2$ and magnetic moment of about $7-8\mu_B$ per Eu atom.
With the crystal lattice parameters of an individual Eu layer in EuRbFe$_4$As$_4$ or EuFe$_2$As$_2$ crystals \cite{Liu_PRB_93_214503} $a=3.9$~$\AA$ and $c/2=6.6$~$\AA$ magnetic moment of Eu atoms corresponds to magnetization $\mu_0 M_s=0.8-0.9$~T, which is above the derived effective magnetization  $\mu_0 M_{eff}$.
This discrepancy indicates that 
(i) either magnetic moment of Eu ions in our samples is about $6.3<7\mu_B$ and the out-of-plane anisotropy field $2K_u/M_s=0$,  
or (ii) magnetic moment of Eu ions does correspond to $\mu_0 M_s=0.8-0.9$~T, and the difference is attributed to positive anisotropy field $2K_u/M_s=\mu_0 M_s-\mu_0 M_{eff}\approx0.1-0.2$~T.
Both scenarios imply that the ab crystal planes are not easy magnetization planes: either magnetocrystalline anisotropy is insignificant, or it corresponds to easy magnetization axis (c-axis) with the anisotropy field of about 0.1-0.2~T.


The dependence of the spin ordering in in $ab$-planes on magnetic field can be found simply by minimizing the free energy Eq.~\ref{E_gen} with predefined $\theta_i=\pi/2$.
However, minimization of such free energy is analytically and numerically challenging.
Analytically such were carried out in Refs.~\cite{Johnston_PRB_91_064427,Johnston_PRB_96_104405,Johnston_PRB_99_214438} by considering a priory conservation of the helical wave-vector.
Direct numerical iterative search of the minimum of free energy is obstructed by a rather large computational size of the problem: for a system larger than 5 spins it is practically impossible find minimums with reasonable precision in orientation $\phi_i$ and reasonable step-size in magnetic field in adequate time. 
In Ref.~\cite{Robinson_PRB_2_2642} minimization was carried-out with the general Newton's method. 
By attempting to apply the same approach we have found that the Newton's method is inconveniently sensitive to initial conditions and parameters of the numerical scheme.

In this work we employ the Nelder-Mead simplex numerical algorithm for search of local minima of the free energy Eq.~\ref{E_gen}.
Typically, we consider a chain of up to a 40 macrospins (typically, from $4N_h$ for large helicity to $10N_h$ for small helicity) and employ periodic boundary conditions.
We use the leap-frog numerical scheme while sweeping the magnetic field up from 0 to $1.3H^x_S$ and down from $1.3H^x_S$ to 0, and calculate the free energy at each field step and each field-sweep direction.
The resulting spin configuration is selected by the absolute minimum of free energy at a given field. 
We find that this scheme is rather robust towards a choice of initial conditions.
Yet, employment of the leap-from scheme and initial helix or fan conditions\cite{Johnston_PRB_96_104405} speeds up the convergence considerably.

\subsection{Numerical results}

In calculations we predefine the saturation field of the spin-system is $\mu_0 H^x_S=1.25$~T and compute $M(H)$ for different helicities $N_h$ in accordance with Eqs.~\ref{H_e}~and~\ref{H_x}.
Calculation results are shown in Fig.~\ref{Th}.
First, we recover numerically basic results for 1D helical spin structures and for helix-to-fan transition, which are fully consistent with previous reports \cite{Robinson_PRB_2_2642,Carazza_ZPhysB_84_301,Johnston_PRB_96_104405} (see Fig.~\ref{Th}a).
For instance, we confirm validity of Eqs.~\ref{H_e}~and~\ref{H_x}: numerically we obtain the predefined saturation field.
We find that for any $N_h>4.5$, which corresponds to $H_{z1}<0$ and $H_{z2}>0$, the dependence $M(H)$ is stepwise.
At magnetic field of magnetization jump $H_{sf}$ the helix-to-fan spin-flop phase transition of the first order occurs.
The helicity at the phase transition is preserved: both the helix and the fan phases conserve periodicity with the wave-length $N_h d_{Eu}$ along the z-axis, where $d_{Eu}$ is the lattice spacing between Eu layers.
At $N_h\gg 1$ the field of the spin-flop transition $H_{sf}$ approaches $H^x_S/2$, and the jump in magnetization $\Delta M_{sf}$ approaches about $0.9M_s$ (see Fig.~\ref{Th}b).
For any $N_h<4.5$, which corresponds to $H_{z1}>0$ and $H_{z2}>0$, the dependence $M(H)$ is continuous (see Fig.~\ref{Th}a), which corresponds to the second-order helix-to-fan transition, and the helicity is not preserved.

Magnetization curves $M(H)$ in Fig.~\ref{MH}h show the stepwise behaviour, which implies that helicity for studied samples is $N_h>4.5$.
According to Fig.~\ref{Th}a,b, both the field of spin-flop transition and the magnetization jump at spin-flop transition show dependence on helicity.
The helicity can be specified by considering exact $H_{sf}$ field and $\Delta M_{sf}$.
In average, the metamagnetic transition in Fig.~\ref{MH}h occurs in a wide range of $H_{sf}/H^x_S\approx 0.45-0.5$, which covers almost the entire range.
However, the change in magnetization in Fig.~\ref{MH}h is $\Delta M/M_s=0.44-0.48$ for all samples.
According to Fig.~\ref{Th}b, this value of  $\Delta M$ corresponds to helicity $N_h\approx 4.8$ for both EuRbFe$_4$As$_4$ and EuFe$_2$As$_2$ compounds with $\mu_0 H_{z1}=-0.59$~T and $\mu_0 H_{z1}=0.57$~T.

\section{Conclusion}

Summarizing, in this work we have synthesized and studied a series of single crystal ferromagnetic superconductors and confirm their crystal structure and composition with XRD, EDX, EBSD and transport measurements.
Structural and composition analysis show that the synthesized crystals consist of EuRbFe$_4$As$_4$ and EuFe$_2$As$_2$ phases.
Magnetization measurements show that the synthesized crystals demonstrate properties of EuRbFe$_4$As$_4$ superconducting ferromagnetic phase and EuFe$_2$As$_2$ nonsuperconducting ferromagnetic phase.
Despite a different composition and magnetic behaviour all studied samples have shown a strikingly similar ferromagnetic properties with identical transition magnetic fields.
This implies that ferromagnetic properties of Eu layers in EuRbFe$_4$As$_4$ and in EuFe$_2$As$_2$ are practically identical at least at temperatures far below the superconducting and magnetic transition temperatures.

Next, we consider helical spin ordering in EuRbFeAs single crystals.
The difference in effective magnetization, provided by saturation fields of helical spin chains, and the nominal saturation magnetization suggest that EuRbFe$_4$As$_4$ and EuFe$_2$As$_2$ single crystals do not shown easy-ab-plane magnetocrystalline anisotropy.
By comparing experimental magnetization dependence on magnetic field with numerical results we find that the observed step-like magnetization curves manifest the helix-to-fan phase transition for the helical spin systems with helicity $N_h\approx 4.8$.

As a final remark we note that within the macrospin approximation the helix-to-fan transition is the first order phase transition where the magnetization changes abruptly (see theoretical curves in Fig.~\ref{Th}a).
However, magnetization measurements shows that the helix-to-fan transition is continuous and occurs within the range of 0.2~T (Fig.~\ref{MH}h and the gray area in Fig.~\ref{Th}a).
This $M(H)$ dependence is attributed to the absence of the easy-plane anisotropy in Eu layers.
Upon approach to the helix-to-fan transition individual Eu layers demagnetize and macrospin behaviour terminates.
At some field after the helix-to-fan transition is passed effective fields magnetize individual Eu layers and the macrospin behaviour restores.

\section{Acknowledgments}

This work was supported by the Russian Science Foundation and by the Ministry of Science and Higher Education of the Russian Federation.

\bibliography{A_Bib_EuRbFeAs}

\providecommand{\latin}[1]{#1}
\makeatletter
\providecommand{\doi}
  {\begingroup\let\do\@makeother\dospecials
  \catcode`\{=1 \catcode`\}=2 \doi@aux}
\providecommand{\doi@aux}[1]{\endgroup\texttt{#1}}
\makeatother
\providecommand*\mcitethebibliography{\thebibliography}
\csname @ifundefined\endcsname{endmcitethebibliography}
  {\let\endmcitethebibliography\endthebibliography}{}
\begin{mcitethebibliography}{45}
\providecommand*\natexlab[1]{#1}
\providecommand*\mciteSetBstSublistMode[1]{}
\providecommand*\mciteSetBstMaxWidthForm[2]{}
\providecommand*\mciteBstWouldAddEndPuncttrue
  {\def\EndOfBibitem{\unskip.}}
\providecommand*\mciteBstWouldAddEndPunctfalse
  {\let\EndOfBibitem\relax}
\providecommand*\mciteSetBstMidEndSepPunct[3]{}
\providecommand*\mciteSetBstSublistLabelBeginEnd[3]{}
\providecommand*\EndOfBibitem{}
\mciteSetBstSublistMode{f}
\mciteSetBstMaxWidthForm{subitem}{(\alph{mcitesubitemcount})}
\mciteSetBstSublistLabelBeginEnd
  {\mcitemaxwidthsubitemform\space}
  {\relax}
  {\relax}

\bibitem[Ren \latin{et~al.}(2009)Ren, \latin{et~al.}
  others]{Ren_PRL_102_137002}
Ren,~Z., \latin{et~al.}  \emph{Phys. Rev. Lett.} \textbf{2009}, \emph{102},
  137002\relax
\mciteBstWouldAddEndPuncttrue
\mciteSetBstMidEndSepPunct{\mcitedefaultmidpunct}
{\mcitedefaultendpunct}{\mcitedefaultseppunct}\relax
\EndOfBibitem
\bibitem[Cao \latin{et~al.}(2011)Cao, Xu, Ren, Jiang, Feng, and
  Xu]{Cao_JPCM_23_464204}
Cao,~G.; Xu,~S.; Ren,~Z.; Jiang,~S.; Feng,~C.; Xu,~Z. \emph{Journal of Physics:
  Condensed Matter} \textbf{2011}, \emph{23}, 464204\relax
\mciteBstWouldAddEndPuncttrue
\mciteSetBstMidEndSepPunct{\mcitedefaultmidpunct}
{\mcitedefaultendpunct}{\mcitedefaultseppunct}\relax
\EndOfBibitem
\bibitem[Jeevan \latin{et~al.}(2011)Jeevan, \latin{et~al.}
  others]{Jeevan_PRB_83_054511}
Jeevan,~H.~S., \latin{et~al.}  \emph{Phys. Rev. B} \textbf{2011}, \emph{83},
  054511\relax
\mciteBstWouldAddEndPuncttrue
\mciteSetBstMidEndSepPunct{\mcitedefaultmidpunct}
{\mcitedefaultendpunct}{\mcitedefaultseppunct}\relax
\EndOfBibitem
\bibitem[Nandi \latin{et~al.}(2014)Nandi, \latin{et~al.}
  others]{Nandi_PRB_89_014512}
Nandi,~S., \latin{et~al.}  \emph{Phys. Rev. B} \textbf{2014}, \emph{89},
  014512\relax
\mciteBstWouldAddEndPuncttrue
\mciteSetBstMidEndSepPunct{\mcitedefaultmidpunct}
{\mcitedefaultendpunct}{\mcitedefaultseppunct}\relax
\EndOfBibitem
\bibitem[Vlasenko \latin{et~al.}(2020)Vlasenko, Pervakov, and
  Gavrilkin]{Vlasenko_SUST_33_084009}
Vlasenko,~V.; Pervakov,~K.; Gavrilkin,~S. \emph{Supercond. Sci. Technol.}
  \textbf{2020}, \emph{33}, 084009\relax
\mciteBstWouldAddEndPuncttrue
\mciteSetBstMidEndSepPunct{\mcitedefaultmidpunct}
{\mcitedefaultendpunct}{\mcitedefaultseppunct}\relax
\EndOfBibitem
\bibitem[Liu \latin{et~al.}(2016)Liu, Liu, Tang, Jiang, Wang, Ablimit, Jiao,
  Tao, Feng, Xu, and Cao]{Liu_PRB_93_214503}
Liu,~Y.; Liu,~Y.-B.; Tang,~Z.-T.; Jiang,~H.; Wang,~Z.-C.; Ablimit,~A.;
  Jiao,~W.-H.; Tao,~Q.; Feng,~C.-M.; Xu,~Z.-A.; Cao,~G.-H. \emph{Phys. Rev. B}
  \textbf{2016}, \emph{93}, 214503\relax
\mciteBstWouldAddEndPuncttrue
\mciteSetBstMidEndSepPunct{\mcitedefaultmidpunct}
{\mcitedefaultendpunct}{\mcitedefaultseppunct}\relax
\EndOfBibitem
\bibitem[Liu \latin{et~al.}(2017)Liu, Liu, Yu, Tao, Feng, and
  Cao]{Liu_PRB_96_224510}
Liu,~Y.; Liu,~Y.-B.; Yu,~Y.-L.; Tao,~Q.; Feng,~C.-M.; Cao,~G.-H. \emph{Phys.
  Rev. B} \textbf{2017}, \emph{96}, 224510\relax
\mciteBstWouldAddEndPuncttrue
\mciteSetBstMidEndSepPunct{\mcitedefaultmidpunct}
{\mcitedefaultendpunct}{\mcitedefaultseppunct}\relax
\EndOfBibitem
\bibitem[Smylie \latin{et~al.}(2018)Smylie, Willa, Bao, Ryan, Islam, Claus,
  Simsek, Diao, Rydh, Koshelev, Kwok, Chung, Kanatzidis, and
  Welp]{Smylie_PRB_98_104503}
Smylie,~M.~P.; Willa,~K.; Bao,~J.-K.; Ryan,~K.; Islam,~Z.; Claus,~H.;
  Simsek,~Y.; Diao,~Z.; Rydh,~A.; Koshelev,~A.~E.; Kwok,~W.-K.; Chung,~D.~Y.;
  Kanatzidis,~M.~G.; Welp,~U. \emph{Phys. Rev. B} \textbf{2018}, \emph{98},
  014503\relax
\mciteBstWouldAddEndPuncttrue
\mciteSetBstMidEndSepPunct{\mcitedefaultmidpunct}
{\mcitedefaultendpunct}{\mcitedefaultseppunct}\relax
\EndOfBibitem
\bibitem[Iida \latin{et~al.}(2019)Iida, Nagai, Ishida, Ishikado, Murai,
  Christianson, Yoshida, Inamura, Nakamura, Nakao, Munakata, Kagerbauer,
  Eisterer, Kawashima, Yoshida, Eisaki, and Iyo]{Iida_PRB_100_014506}
Iida,~K. \latin{et~al.}  \emph{Phys. Rev. B} \textbf{2019}, \emph{100},
  014506\relax
\mciteBstWouldAddEndPuncttrue
\mciteSetBstMidEndSepPunct{\mcitedefaultmidpunct}
{\mcitedefaultendpunct}{\mcitedefaultseppunct}\relax
\EndOfBibitem
\bibitem[Ghigo \latin{et~al.}(2019)Ghigo, \latin{et~al.}
  others]{Ghigo_PRR_1_033110}
Ghigo,~G., \latin{et~al.}  \emph{Phys. Rev. Res.} \textbf{2019}, \emph{1},
  033110\relax
\mciteBstWouldAddEndPuncttrue
\mciteSetBstMidEndSepPunct{\mcitedefaultmidpunct}
{\mcitedefaultendpunct}{\mcitedefaultseppunct}\relax
\EndOfBibitem
\bibitem[Kim \latin{et~al.}(2021)Kim, Pervakov, Evtushinsky, Jung, Poelchen,
  Kummer, Vlasenko, Sadakov, Usoltsev, Pudalov, Roditchev, Stolyarov, Vyalikh,
  Borisov, Valent, Ernst, Eremeev, and Chulkov]{Kim_PRB}
Kim,~T.~K. \latin{et~al.}  \emph{Phys. Rev. B} \textbf{2021}, \emph{When
  superconductivity does not fear magnetism}, 0\relax
\mciteBstWouldAddEndPuncttrue
\mciteSetBstMidEndSepPunct{\mcitedefaultmidpunct}
{\mcitedefaultendpunct}{\mcitedefaultseppunct}\relax
\EndOfBibitem
\bibitem[Stolyarov \latin{et~al.}(2020)Stolyarov, Pervakov, Astrakhantseva,
  Golovchanskiy, Vyalikh, Kim, Eremeev, Vlasenko, Pudalov, Golubov, Chulkov,
  and Roditchev]{Stolyarov_JPCL_11_9393}
Stolyarov,~V.~S.; Pervakov,~K.~S.; Astrakhantseva,~A.~S.; Golovchanskiy,~I.~A.;
  Vyalikh,~D.~V.; Kim,~T.~K.; Eremeev,~S.~V.; Vlasenko,~V.~A.; Pudalov,~V.~M.;
  Golubov,~A.~A.; Chulkov,~E.~V.; Roditchev,~D. \emph{J. Phys. Chem. Lett.}
  \textbf{2020}, \emph{11}, 9393\relax
\mciteBstWouldAddEndPuncttrue
\mciteSetBstMidEndSepPunct{\mcitedefaultmidpunct}
{\mcitedefaultendpunct}{\mcitedefaultseppunct}\relax
\EndOfBibitem
\bibitem[Jiang \latin{et~al.}(2009)Jiang, Luo, Ren, Zhu, Wang, Xu, Tao, Cao,
  and Xu]{Jiang_NJP_11_025007}
Jiang,~S.; Luo,~Y.; Ren,~Z.; Zhu,~Z.; Wang,~C.; Xu,~X.; Tao,~Q.; Cao,~G.;
  Xu,~Z. \emph{New J. Phys.} \textbf{2009}, \emph{11}, 025007\relax
\mciteBstWouldAddEndPuncttrue
\mciteSetBstMidEndSepPunct{\mcitedefaultmidpunct}
{\mcitedefaultendpunct}{\mcitedefaultseppunct}\relax
\EndOfBibitem
\bibitem[Stolyarov \latin{et~al.}(2018)Stolyarov, \latin{et~al.}
  others]{Stolyarov_SciAdv_4_eaat1061}
Stolyarov,~V.~S., \latin{et~al.}  \emph{Sci. Adv.} \textbf{2018}, \emph{4},
  eaat1061\relax
\mciteBstWouldAddEndPuncttrue
\mciteSetBstMidEndSepPunct{\mcitedefaultmidpunct}
{\mcitedefaultendpunct}{\mcitedefaultseppunct}\relax
\EndOfBibitem
\bibitem[Grebenchuk \latin{et~al.}(2020)Grebenchuk, \latin{et~al.}
  others]{Grebenchuk_PRB_102_144501}
Grebenchuk,~S., \latin{et~al.}  \emph{Phys. Rev. B} \textbf{2020}, \emph{102},
  144501\relax
\mciteBstWouldAddEndPuncttrue
\mciteSetBstMidEndSepPunct{\mcitedefaultmidpunct}
{\mcitedefaultendpunct}{\mcitedefaultseppunct}\relax
\EndOfBibitem
\bibitem[Koshelev(2019)]{Koshelev_PRB_100_224503}
Koshelev,~A.~E. \emph{Phys. Rev. B} \textbf{2019}, \emph{100}, 224503\relax
\mciteBstWouldAddEndPuncttrue
\mciteSetBstMidEndSepPunct{\mcitedefaultmidpunct}
{\mcitedefaultendpunct}{\mcitedefaultseppunct}\relax
\EndOfBibitem
\bibitem[Xiao \latin{et~al.}(2009)Xiao, Su, Meven, Mittal, Kumar, Chatterji,
  Price, Persson, Kumar, Dhar, Thamizhavel, and Brueckel]{Xiao_PRB_80_174424}
Xiao,~Y.; Su,~Y.; Meven,~M.; Mittal,~R.; Kumar,~C. M.~N.; Chatterji,~T.;
  Price,~S.; Persson,~J.; Kumar,~N.; Dhar,~S.~K.; Thamizhavel,~A.; Brueckel,~T.
  \emph{Phys. Rev. B} \textbf{2009}, \emph{80}, 174424\relax
\mciteBstWouldAddEndPuncttrue
\mciteSetBstMidEndSepPunct{\mcitedefaultmidpunct}
{\mcitedefaultendpunct}{\mcitedefaultseppunct}\relax
\EndOfBibitem
\bibitem[Collomb \latin{et~al.}(2021)Collomb, Bending, Koshelev, Smylie,
  Farrar, Bao, Chung, Kanatzidis, Kwok, and Welp]{Collomb_PRL_126_157001}
Collomb,~D.; Bending,~S.~J.; Koshelev,~A.~E.; Smylie,~M.~P.; Farrar,~L.;
  Bao,~J.-K.; Chung,~D.~Y.; Kanatzidis,~M.~G.; Kwok,~W.-K.; Welp,~U.
  \emph{Phys. Rev. Lett.} \textbf{2021}, \emph{126}, 157001\relax
\mciteBstWouldAddEndPuncttrue
\mciteSetBstMidEndSepPunct{\mcitedefaultmidpunct}
{\mcitedefaultendpunct}{\mcitedefaultseppunct}\relax
\EndOfBibitem
\bibitem[Hemmida \latin{et~al.}(2021)Hemmida, Winterhalter-Stocker, Ehlers, von
  Nidda, Yao, Bannies, Rienks, Kurleto, Felser, B{\"u}chner, Fink, Gorol,
  F{\"o}rster, Arsenijevic, Fritsch, and Gegenwart]{Hemmida_PRB_103_195112}
Hemmida,~M. \latin{et~al.}  \emph{Phys. Rev. B} \textbf{2021}, \emph{103},
  195112\relax
\mciteBstWouldAddEndPuncttrue
\mciteSetBstMidEndSepPunct{\mcitedefaultmidpunct}
{\mcitedefaultendpunct}{\mcitedefaultseppunct}\relax
\EndOfBibitem
\bibitem[Nagamiya(1968)]{Nagamiya_SSP_20_305}
Nagamiya,~T. \emph{Solid State Physics} \textbf{1968}, \emph{20}, 305\relax
\mciteBstWouldAddEndPuncttrue
\mciteSetBstMidEndSepPunct{\mcitedefaultmidpunct}
{\mcitedefaultendpunct}{\mcitedefaultseppunct}\relax
\EndOfBibitem
\bibitem[Robinson and Erdos(1970)Robinson, and Erdos]{Robinson_PRB_2_2642}
Robinson,~J.~M.; Erdos,~P. \emph{Phys. Rev. B} \textbf{1970}, \emph{2},
  2642\relax
\mciteBstWouldAddEndPuncttrue
\mciteSetBstMidEndSepPunct{\mcitedefaultmidpunct}
{\mcitedefaultendpunct}{\mcitedefaultseppunct}\relax
\EndOfBibitem
\bibitem[Johnston(2012)]{Johnston_PRL_109_077201}
Johnston,~D.~C. \emph{Phys. Rev. Lett.} \textbf{2012}, \emph{109}, 077201\relax
\mciteBstWouldAddEndPuncttrue
\mciteSetBstMidEndSepPunct{\mcitedefaultmidpunct}
{\mcitedefaultendpunct}{\mcitedefaultseppunct}\relax
\EndOfBibitem
\bibitem[Johnston(2015)]{Johnston_PRB_91_064427}
Johnston,~D. \emph{Phys. Rev. B} \textbf{2015}, \emph{91}, 064427\relax
\mciteBstWouldAddEndPuncttrue
\mciteSetBstMidEndSepPunct{\mcitedefaultmidpunct}
{\mcitedefaultendpunct}{\mcitedefaultseppunct}\relax
\EndOfBibitem
\bibitem[Pervakov \latin{et~al.}(2013)Pervakov, Vlasenko, Khlybov, Zaleski,
  Pudalov, and Eltsev]{Pervakov_SUST_26_015008}
Pervakov,~K.~S.; Vlasenko,~V.~A.; Khlybov,~E.~P.; Zaleski,~A.; Pudalov,~V.~M.;
  Eltsev,~Y.~F. \emph{Supercond. Sci. Technol.} \textbf{2013}, \emph{26},
  015008\relax
\mciteBstWouldAddEndPuncttrue
\mciteSetBstMidEndSepPunct{\mcitedefaultmidpunct}
{\mcitedefaultendpunct}{\mcitedefaultseppunct}\relax
\EndOfBibitem
\bibitem[Kawashima \latin{et~al.}(2016)Kawashima, Kinjo, Nishio, Ishida,
  Fujihisa, Gotoh, Kihou, Eisaki, Yoshida, and Iyo]{Kawashima_JPSJ_85_064710}
Kawashima,~K.; Kinjo,~T.; Nishio,~T.; Ishida,~S.; Fujihisa,~H.; Gotoh,~Y.;
  Kihou,~K.; Eisaki,~H.; Yoshida,~Y.; Iyo,~A. \emph{J. Phys. Soc. Jpn.}
  \textbf{2016}, \emph{85}, 064710\relax
\mciteBstWouldAddEndPuncttrue
\mciteSetBstMidEndSepPunct{\mcitedefaultmidpunct}
{\mcitedefaultendpunct}{\mcitedefaultseppunct}\relax
\EndOfBibitem
\bibitem[Jeevan \latin{et~al.}(2008)Jeevan, Hossain, Kasinathan, Rosner,
  Geibel, and Gegenwart]{Jeevan_PRB_78_052502}
Jeevan,~H.~S.; Hossain,~Z.; Kasinathan,~D.; Rosner,~H.; Geibel,~C.;
  Gegenwart,~P. \emph{Phys. Rev. B} \textbf{2008}, \emph{78}, 052502\relax
\mciteBstWouldAddEndPuncttrue
\mciteSetBstMidEndSepPunct{\mcitedefaultmidpunct}
{\mcitedefaultendpunct}{\mcitedefaultseppunct}\relax
\EndOfBibitem
\bibitem[Ren \latin{et~al.}(2008)Ren, Zhu, Jiang, Xu, Tao, Wang, Feng, Cao, and
  Xu]{Ren_PRB_78_052501}
Ren,~Z.; Zhu,~Z.; Jiang,~S.; Xu,~X.; Tao,~Q.; Wang,~C.; Feng,~C.; Cao,~G.;
  Xu,~Z. \emph{Phys. Rev. B} \textbf{2008}, \emph{78}, 052501\relax
\mciteBstWouldAddEndPuncttrue
\mciteSetBstMidEndSepPunct{\mcitedefaultmidpunct}
{\mcitedefaultendpunct}{\mcitedefaultseppunct}\relax
\EndOfBibitem
\bibitem[Golubov \latin{et~al.}(2011)Golubov, Dolgov, Boris, Charnukha, Sun,
  Lin, Shevchun, Korobenko, Trunin, and Zverev]{Golubov_JETPLet_94_333}
Golubov,~A.~A.; Dolgov,~O.~V.; Boris,~A.~V.; Charnukha,~A.; Sun,~D.~L.;
  Lin,~C.~T.; Shevchun,~A.~F.; Korobenko,~A.~V.; Trunin,~M.~R.; Zverev,~V.~N.
  \emph{JETP Letters} \textbf{2011}, \emph{94}, 333\relax
\mciteBstWouldAddEndPuncttrue
\mciteSetBstMidEndSepPunct{\mcitedefaultmidpunct}
{\mcitedefaultendpunct}{\mcitedefaultseppunct}\relax
\EndOfBibitem
\bibitem[Bean(1964)]{BEAN_RMP_36_31}
Bean,~C.~P. \emph{Rev. Mod. Phys.} \textbf{1964}, \emph{36}, 31\relax
\mciteBstWouldAddEndPuncttrue
\mciteSetBstMidEndSepPunct{\mcitedefaultmidpunct}
{\mcitedefaultendpunct}{\mcitedefaultseppunct}\relax
\EndOfBibitem
\bibitem[Norris(1970)]{Norris_JPDAP_3_489}
Norris,~W.~T. \emph{J. Phys. D: Appl. Phys.} \textbf{1970}, \emph{3}, 489\relax
\mciteBstWouldAddEndPuncttrue
\mciteSetBstMidEndSepPunct{\mcitedefaultmidpunct}
{\mcitedefaultendpunct}{\mcitedefaultseppunct}\relax
\EndOfBibitem
\bibitem[Chen and Goldfarb(1989)Chen, and Goldfarb]{Chen_JAP_66_2489}
Chen,~D.; Goldfarb,~R.~B. \emph{J. Appl. Phys.} \textbf{1989}, \emph{66},
  2489\relax
\mciteBstWouldAddEndPuncttrue
\mciteSetBstMidEndSepPunct{\mcitedefaultmidpunct}
{\mcitedefaultendpunct}{\mcitedefaultseppunct}\relax
\EndOfBibitem
\bibitem[Hwang \latin{et~al.}(2009)Hwang, Choi, Cho, Ketterson, and
  Tsai]{Hwang_JAP_105_07A946}
Hwang,~Y.; Choi,~J.; Cho,~S.; Ketterson,~J.~B.; Tsai,~C.-C. \emph{J. Appl.
  Phys.} \textbf{2009}, \emph{105}, 07A946\relax
\mciteBstWouldAddEndPuncttrue
\mciteSetBstMidEndSepPunct{\mcitedefaultmidpunct}
{\mcitedefaultendpunct}{\mcitedefaultseppunct}\relax
\EndOfBibitem
\bibitem[Yang \latin{et~al.}(2020)Yang, Kang, Diao, Karigerasi, Shoemaker,
  Schleife, and Cahil]{Yang_PRB_102_064415}
Yang,~K.; Kang,~K.; Diao,~Z.; Karigerasi,~M.~H.; Shoemaker,~D.~P.;
  Schleife,~A.; Cahil,~D.~G. \emph{Phys. Rev. B} \textbf{2020}, \emph{102},
  064415\relax
\mciteBstWouldAddEndPuncttrue
\mciteSetBstMidEndSepPunct{\mcitedefaultmidpunct}
{\mcitedefaultendpunct}{\mcitedefaultseppunct}\relax
\EndOfBibitem
\bibitem[Nagamiya \latin{et~al.}(1962)Nagamiya, Nagata, and
  Kitano]{Nagamiya_ProgTheorPhys_27_1253}
Nagamiya,~T.; Nagata,~K.; Kitano,~Y. \emph{Progress of Theoretical Physics}
  \textbf{1962}, \emph{27}, 1253\relax
\mciteBstWouldAddEndPuncttrue
\mciteSetBstMidEndSepPunct{\mcitedefaultmidpunct}
{\mcitedefaultendpunct}{\mcitedefaultseppunct}\relax
\EndOfBibitem
\bibitem[Carazza \latin{et~al.}(1991)Carazza, Rastelli, and
  Tassi]{Carazza_ZPhysB_84_301}
Carazza,~B.; Rastelli,~E.; Tassi,~A. \emph{Z. Physik B - Condensed Matter}
  \textbf{1991}, \emph{84}, 301\relax
\mciteBstWouldAddEndPuncttrue
\mciteSetBstMidEndSepPunct{\mcitedefaultmidpunct}
{\mcitedefaultendpunct}{\mcitedefaultseppunct}\relax
\EndOfBibitem
\bibitem[Johnston(2017)]{Johnston_PRB_96_104405}
Johnston,~D. \emph{Phys. Rev. B} \textbf{2017}, \emph{96}, 104405\relax
\mciteBstWouldAddEndPuncttrue
\mciteSetBstMidEndSepPunct{\mcitedefaultmidpunct}
{\mcitedefaultendpunct}{\mcitedefaultseppunct}\relax
\EndOfBibitem
\bibitem[Reehuis \latin{et~al.}(1992)Reehuis, Jeitschko, Moller, and
  Brown]{Reehuis_JPhysChemSolids_53_687}
Reehuis,~M.; Jeitschko,~W.; Moller,~M.~H.; Brown,~P.~J. \emph{Journal of
  Physics and Chemistry of Solids} \textbf{1992}, \emph{53}, 687\relax
\mciteBstWouldAddEndPuncttrue
\mciteSetBstMidEndSepPunct{\mcitedefaultmidpunct}
{\mcitedefaultendpunct}{\mcitedefaultseppunct}\relax
\EndOfBibitem
\bibitem[Sangeetha \latin{et~al.}(2018)Sangeetha, Anand, Cuervo-Reyes, Smetana,
  Mudring, and Johnston]{Sangeetha_PRB_97_144403}
Sangeetha,~N.~S.; Anand,~V.~K.; Cuervo-Reyes,~E.; Smetana,~V.; Mudring,~A.-V.;
  Johnston,~D.~C. \emph{Phys. Rev. B} \textbf{2018}, \emph{97}, 144403\relax
\mciteBstWouldAddEndPuncttrue
\mciteSetBstMidEndSepPunct{\mcitedefaultmidpunct}
{\mcitedefaultendpunct}{\mcitedefaultseppunct}\relax
\EndOfBibitem
\bibitem[Sangeetha \latin{et~al.}(2016)Sangeetha, Cuervo-Reyes, Pandey, and
  Johnston]{Sangeetha_PRB_94_014422}
Sangeetha,~N.~S.; Cuervo-Reyes,~E.; Pandey,~A.; Johnston,~D.~C. \emph{Phys.
  Rev. B} \textbf{2016}, \emph{94}, 014422\relax
\mciteBstWouldAddEndPuncttrue
\mciteSetBstMidEndSepPunct{\mcitedefaultmidpunct}
{\mcitedefaultendpunct}{\mcitedefaultseppunct}\relax
\EndOfBibitem
\bibitem[Fabreges \latin{et~al.}(2016)Fabreges, Gukasov, Bonville, Maurya,
  Thamizhavel, and Dhar]{Fabreges_PRB_93_214414}
Fabreges,~X.; Gukasov,~A.; Bonville,~P.; Maurya,~A.; Thamizhavel,~A.;
  Dhar,~S.~K. \emph{Phys. Rev. B} \textbf{2016}, \emph{93}, 214414\relax
\mciteBstWouldAddEndPuncttrue
\mciteSetBstMidEndSepPunct{\mcitedefaultmidpunct}
{\mcitedefaultendpunct}{\mcitedefaultseppunct}\relax
\EndOfBibitem
\bibitem[Enz(1961)]{Enz_JAP_32_S22}
Enz,~U. \emph{J. Appl. Phys.} \textbf{1961}, \emph{32}, S22\relax
\mciteBstWouldAddEndPuncttrue
\mciteSetBstMidEndSepPunct{\mcitedefaultmidpunct}
{\mcitedefaultendpunct}{\mcitedefaultseppunct}\relax
\EndOfBibitem
\bibitem[Kitano and Nagamiya(1964)Kitano, and
  Nagamiya]{Kitano_ProgTheorPhys_31_1}
Kitano,~Y.; Nagamiya,~T. \emph{Progress of Theoretical Physics} \textbf{1964},
  \emph{31}, 1\relax
\mciteBstWouldAddEndPuncttrue
\mciteSetBstMidEndSepPunct{\mcitedefaultmidpunct}
{\mcitedefaultendpunct}{\mcitedefaultseppunct}\relax
\EndOfBibitem
\bibitem[Johnston(2019)]{Johnston_PRB_99_214438}
Johnston,~D. \emph{Phys. Rev. B} \textbf{2019}, \emph{99}, 214438\relax
\mciteBstWouldAddEndPuncttrue
\mciteSetBstMidEndSepPunct{\mcitedefaultmidpunct}
{\mcitedefaultendpunct}{\mcitedefaultseppunct}\relax
\EndOfBibitem
\bibitem[Ishida \latin{et~al.}(2019)Ishida, Iyo, Ogin, Eisaki, Takeshita,
  Kawashima, Yanagisawa, Kobayashi, Kimoto, Abe, Imai, ichi Shimoyama, and
  Eisterer]{Ishida_npjQM_4_27}
Ishida,~S.; Iyo,~A.; Ogin,~H.; Eisaki,~H.; Takeshita,~N.; Kawashima,~K.;
  Yanagisawa,~K.; Kobayashi,~Y.; Kimoto,~K.; Abe,~H.; Imai,~M.; ichi
  Shimoyama,~J.; Eisterer,~M. \emph{npj Quantum Materials} \textbf{2019},
  \emph{4}, 27\relax
\mciteBstWouldAddEndPuncttrue
\mciteSetBstMidEndSepPunct{\mcitedefaultmidpunct}
{\mcitedefaultendpunct}{\mcitedefaultseppunct}\relax
\EndOfBibitem
\end{mcitethebibliography}

\end{document}